\documentclass[]{tCPH2e}
\usepackage{color}

\begin{document}
\doi{10.1080/0010751YYxxxxxxxx}
 \issn{1366-5812}
\issnp{0010-7514}

\jvol{00} \jnum{00} \jyear{2014}

\markboth{Taylor \& Francis and I.T. Consultant}{Contemporary Physics}

\title{Micro and Macro Benefits of Random Investments in Financial Markets}

\author{A.E.Biondo$^{a}$, A.Pluchino$^{b}$, A.Rapisarda$^{b}$$^{\ast}$\thanks{$^\ast$Corresponding author. Email: andrea.rapisarda@ct.infn.unict.it \vspace{6pt}} 
\\
\vspace{6pt} 
$^{a}${\em{Dipartimento di Economia e Impresa - Universit\'a di Catania, Corso Italia 55, 95129 Catania, Italy}};
$^{b}${\em{Dipartimento di Fisica e Astronomia, Universit\'a di Catania, and INFN Sezione Catania, \mbox{Via S. Sofia 64,} 95123 Catania, Italy}}; \\
\vspace{6pt}}

\maketitle

\begin{abstract}
{
In this paper, making use of recent statistical physics techniques and models, we address the specific role of randomness in financial markets, both at the micro and the macro level. In particular, we review some recent results obtained about the effectiveness of random strategies of investment, compared with some of the most used trading strategies for forecasting the behavior of real financial indexes. We also push forward our analysis by means of a Self-Organized Criticality model, able to simulate financial avalanches in trading communities with different network topologies, where a Pareto-like power law behavior of wealth spontaneously emerges. In this context, we present new findings and suggestions for policies based on the effects that random strategies can have in terms of reduction of dangerous financial extreme events, i.e. bubbles and crashes.

\bigskip

\bigskip
\bigskip

}

\end{abstract}

\section{Introduction}

In the last decades physics and in particular statistical mechanics have influenced in a significant way other fields. Economics, financial markets  and social sciences are just   a few examples \cite{MantegnaStanley,McCauley,Bouchaud,Bass,Sornette-libro,Galam,Stauffer,Helbing1,Castellano}. 
On the other hand, the idea that natural sciences and physics can provide useful tools for deeper analysis and more analytic results is not a novelty in the economics literature, as Marshall recalled in 1885: \textit{At last the speculations of biology made a great stride forwards: its discoveries fascinated the attention of all men as those of physics had done in earlier years. The moral and historical sciences of the day have in consequence changed their tone, and Economics has shared in the general movement. } \cite{hart}.
\\
The aggregate macroeconomic scenario, with its complex interaction between agents and institutions, has been studied and described by several points of view. In particular, as a matter of methodology, one of the most challenging questions is whether to consider the aggregate system either as a simple sum of single individuals (with their singular properties whose collective interaction would then represent a sort of average of them), or as an emergent organism with its own properties, synchronization, herding, and asymmetric volatile aggregate behaviors, whose qualitative characteristics are very different from the simple sum of individual behaviors. 
\\ 
Such a consciousness questions the possibility of  predicting  the future values of variables in economics, especially in financial markets. Economic systems are influenced by expectations, both present and past: many feedback-influenced systems operate and agents' expectations are self-fulfilled and determine the future dynamics. This inspired much attention to the mechanisms of beliefs formation: Simon \cite{simon} underlined that agents' decisions are based on limited knowledge and thus they have to pay high costs to buy information. This create a \textit{bounded rationality} framework for decision making. In contrast to this view, the definition of perfectly rational agents, implies  that the behavior of agents can be described \textit{presuming} their full rationality, since the market clearing mechanism would immediately drive out non-rational agents  \cite{friedman}. 
What appears here is a dichotomy that is deeply rooted within the epistemological debate, very well known in the economic literature, between methodological individualism and aggregate analysis, i.e. the problem of the micro-foundation of macroeconomics. From this perspective, a multidisciplinary approach may greatly help:  the adoption of powerful techniques from  statistical physics and the use of agent-based models  simulations  in studying socio-economic phenomena may lead to innovative and robust results. 
\\
In this paper we follow this line of research, taking inspiration from useful analogies with physical phenomena where noise plays an important role. 
Actually, noise and randomness are very familiar to physicists. In experiments one usually tries to avoid the effect of both of them, since they can perturb the phenomenon under investigation and mask the laws under scrutiny. However, it is well known that quite often, in particular if present in a  limited amount, they can have an important and constructive role which physicists have  realized for a  long time. 
\\
The use of random numbers to calculate in a quick and efficient way complicated integrals or simulate the behavior of a complicated detector (the so called  Monte Carlo algorithm) was introduced by Ulam, Von Neumann and Metropolis  \cite{MC} in Los Alamos during the second world war, when they were working for the "Manhattan project". Since then,  it has been refined to become a fundamental tool for experimental and theoretical physics, being nowadays a scientific topic in itself  \cite{Binder}. In the 80's, investigating climate dynamics, a very interesting phenomenon called "stochastic resonance", where stochastic fluctuations play a significant role, was discovered  by several authors \cite{Benzi,Nicolis}. They realized that random noise can  amplify a weak period forcing,  giving rise to a resonance able to explain the observed stable periodic climate oscillations. This mechanism has been proven to be very  general and it  has found many successful applications in a large variety of physical systems  both at the classical and quantum level (see ref. \cite{Gammaitoni} for a review). 
\\
But physical phenomena are not the only ones that can benefit from noise and randomness: in fact, it is assumed that the noise produced by the random action of many elementary constituents or by the environment, has also a great influence  in the complex dynamics of living cells, of neurons and of many other biological systems \cite{Thurner1,Thurner2,Caruso1,Mossa}. It is then very likely  that  many other  dynamical systems, including  socio-economic organizations, could share a similar situation.  
\\
In recent years, many physicists have started to investigate the complex dynamics of several phenomena beyond the field of physics. In this respect, new disciplines, namely Econophysics \cite{MantegnaStanley,McCauley,Bouchaud,Bass,Sornette-libro} and  Sociophysics \cite{Galam,Stauffer,Helbing1,Castellano}, have been developed pushing forward the availability of more advanced analytical and statistical tools from physics to socio-economic analysis. 
 \\
Along this line of research, the role of  stochastic noise and, more specifically, that of random strategies in several socio-economic applications, have been investigated to try  to understand their eventual positive and constructive features. For example, it has been studied how random strategies of promotions can help to face the problems raised by the so-called Peter principle in hierarchical organizations \cite{Peter, Pluchino1, Pluchino2} or how randomly selected legislators may improve the efficiency of a public institution like a Parliament \cite{Pluchino3}. Other groups have successfully explored the success of similar stochastic strategies also in minority and Parrondo games \cite{Sornette2}. 
\\
In this respect, it  seems of particular interest  the investigation of the role of random strategies in financial markets.
In 2001 the English psychologist  R. Wiseman explored the potentiality of random investments in a famous experiment, where a five-year old child, playing at random with shares of the London Stock Exchange, managed to contain losses better than a financial trader and an astrologist during one year of turbulent market behaviour \cite{Wiseman}. 
Similar results were obtained also in other studies, by exploiting dartboard or monkeys \cite{Porter,Cass}. 
Stimulated by these findings, in the last years we started to investigate in detail the efficiency of random trading with respect to standard technical strategies, both from an individual point of view \cite{Biondo1,Biondo2} and from a collective perspective \cite{Biondo3}, making use of statistical analysis and agent based simulations.
In this paper we will review these results,  presenting also new intriguing and original findings. 
\\
The paper is organized as follows. 
In Section 2 we present a brief overview of micro and macro level approaches in Economics, with reference to the debate between the individualistic approach of perfectly informed rational agents, linked to the Efficient Market Hypothesis and the aggregate view of the collective approach, linked to the Keynesian tradition.
Section 3 reviews results about micro-level investigations of the effectiveness of random strategies against technical analysis, while
Section 4 extends the analysis to the macro-level and explores the role of randomness in reducing financial crises, represented as avalanches in a social system that self-organizes in a critical state. In particular, we consider large trading communities with different network topologies, and we study how the adopted trading strategies influence both the propagation of information and the personal wealth of the investors.   
Finally, some suggestions about possible convenient policies for the stabilization of financial markets will be illustrated in section 5 together with  conclusive remarks.

\section{Micro and Macro Approaches in Economics}

Quite often, the argument of the predictive capacity of economic crises rises to question the ability of economists to contribute to science. Prediction needs natural laws, in the sense of laws of nature, that can be tested  unambiguously in time, under controlled  conditions. Economic systems cannot correspond to this framework, since they depend on  people who base their choices and behaviors on personal opinions, tastes, attitudes, maybe emotions, which are not necessarily replicable, even in identical surrounding scenarios. Those approaches that consider agents as a set of replicas and pursue the analysis of macroeconomic problems starting from the simple sum of deterministic laws valid for single individuals, present, in our opinion, several descriptive limits. In a microeconomic context, the perfect rationality as the expression of the maximization of the interest of each acting agent, can be satisfactory since every person may have whichever goal and may try to reach it; on the contrary, from a macroeconomic (aggregate) point of view, the composition of society becomes a new identity which cannot be successfully described by the same set of instruments. 
\\ 
The marginal approach focused on  individual analysis and established individual foundations of the economic behavior. This is the core element of the neo-classical school, so called in order to qualify it as descendant from the Fathers of Economic Thought - namely Smith, Ricardo, Malthus, whose philosophical investigation was always referred to the socio-institutional context \cite{smith, ricardo, malthus}. Economists such as Marshall, Edgeworth, Jevons, Walras, Bohm-Bawerk, Menger, Fisher, and Pareto (just to mention a few) built in their contributions \cite{marshall, edgeworth, jevons, walras, bohm-bawerk, menger, fisher, pareto} the microeconomic framework that characterizes the rational individual agent who participates to markets.
\\ 
With the Keynesian revolution \cite{keynes}, for the first time, collective behaviors assumed a completely renewed role, with a well-defined relevance in influencing the dynamics of the entire economy. This approach, on one hand, abandons  the chance to describe exactly what one agent would do by means of her presumed infallible rationality. On the other hand, it looks at the social interaction mechanism, somehow linking to Marxian social classes, not in the sense of contraposition yet, but (now) from a viewpoint of different social groups, with different roles and objectives (such as entrepreneurs and workers). Even without participating in  the philosophical debate about micro- or macro-foundations of macroeconomic analysis, one can nonetheless understand the methodological difference between these two approaches. The key-point is that, differently from the microeconomic point of view, the macroeconomic perspective reveals the existence of emergent qualitative phenomena, generated unavoidably by the interaction among individual agents, whose compliance with the microeconomic point of view is not granted. 
\\ 
Such a methodological issue influences the way each market participant forms expectations for future values of variables and thus implies a feedback mechanism which operates in turn on the markets again. Especially for financial markets, it is true that the microeconomic perspective of the single investor is truly different from the aggregate market behavior, which does not respond to any individual motive, leaving aside any possibility to infer without uncertainty the future evolution. 
\\
In summary, there exist two reference models of expectations in the economic literature, namely the adaptive expectations model and the rational expectation model. The former is founded on a somehow weighted average of past values and observational errors may result in repeatedly mistaken predictions. Instead, the latter assumes that all agents know perfectly all the available information and the model that describes the economy, therefore no systematic mistakes are possible. These approaches have been introduced in, \textit{inter alia}, Arrow and Nerlove \cite{arrow-nerlove}, Friedman \cite{friedman,friedman2}, Phelps \cite{phelps} and Cagan \cite{cagan} (for adaptive expectations), and Muth \cite{muth}, Lucas \cite{lucas} and Sargent-Wallace \cite{sargent-wallace} (for rational expectations). The recent Nobel Laureate Fama \cite{fama} defined financial efficiency depending on the existence of \emph{perfect arbitrage}: the author suggests three forms of market efficiency - namely ``weak'', ``semi-strong'', and ``strong'' - according to the degree of completeness of the informative set. Inefficiency would then imply the existence of opportunities for unexploited profits that traders would immediately try to exploit. Other authors, such as Jensen \cite{jensen} and Malkiel \cite{malkiel}, also link the efficiency of the available information to the  determination of  assets' prices. Then, financial markets participants continuously seek to expand their informative set to choose the best strategy and this results in extreme variability and high volatility. 
\\ 
The so-called Efficient Market Hypothesis (whose main theoretical background is the theory of rational expectations), describes the case of perfectly competitive markets and perfectly rational agents, endowed with all available information, who choose  the best strategies (since otherwise the competitive clearing mechanism would put them out of the market). There is, however, evidence that this interpretation of a fully working \textit{perfect arbitrage} mechanism, without systematic forecasting errors, is not adequate to analyze financial markets: Cutler\emph{et al.} \cite{cutler}, Engle \cite{engle}, Mandelbrot \cite{mandelbrot,mandelbrot2}, Lux \cite{lux2}, and Mantegna and Stanley \cite{MantegnaStanley} just to mention some examples. The reason is quite intuitive: the hypothesis that information is available for everybody is not real. And it is not real even if in its semi-strong or weak versions. Many heterogeneous agent models have been introduced in the field of financial literature in order to describe what happens in true markets with different types of traders, each with different expectations, influencing each other by means of the consequences of their behaviors (some examples are: Brock \cite{brock,brock2}, Brock and Hommes \cite{brock-hommes}, Chiarella \cite{chiarella}, Chiarella and He \cite{chiarella-he}, DeGrauwe \emph{et al.} \cite{degrauwe}, Frankel and Froot \cite{frankel-froot}, Lux \cite{lux}, Wang \cite{wang}, and Zeeman\cite{zeeman}). This approach, namely the ``adaptive belief systems'', tries to apply non-linearity and noise to financial market models in order to represent the highly complex behavior of markets. In this respect, it favors an interdisciplinary approach, based on statistical physics techniques and economic analysis, which - as we will show in the following - can lead to major advances.

\section{Profitability of random investments at a micro level}

In this section we explore the effectiveness of random investments from the point of view of the single trader (micro level), leaving the analysis of the emergent collective behavior of many interacting traders (macro level) to the next section. Thus, we consider here only three non interacting agents $A_i$ $(i=1,2,3)$, which invest daily, for a long time period, in two stock markets by adopting different trading strategies. Our simulations will tell us which strategy is more profitable, both over long and short periods.   

\begin{figure}  
\begin{center}
\epsfig{figure=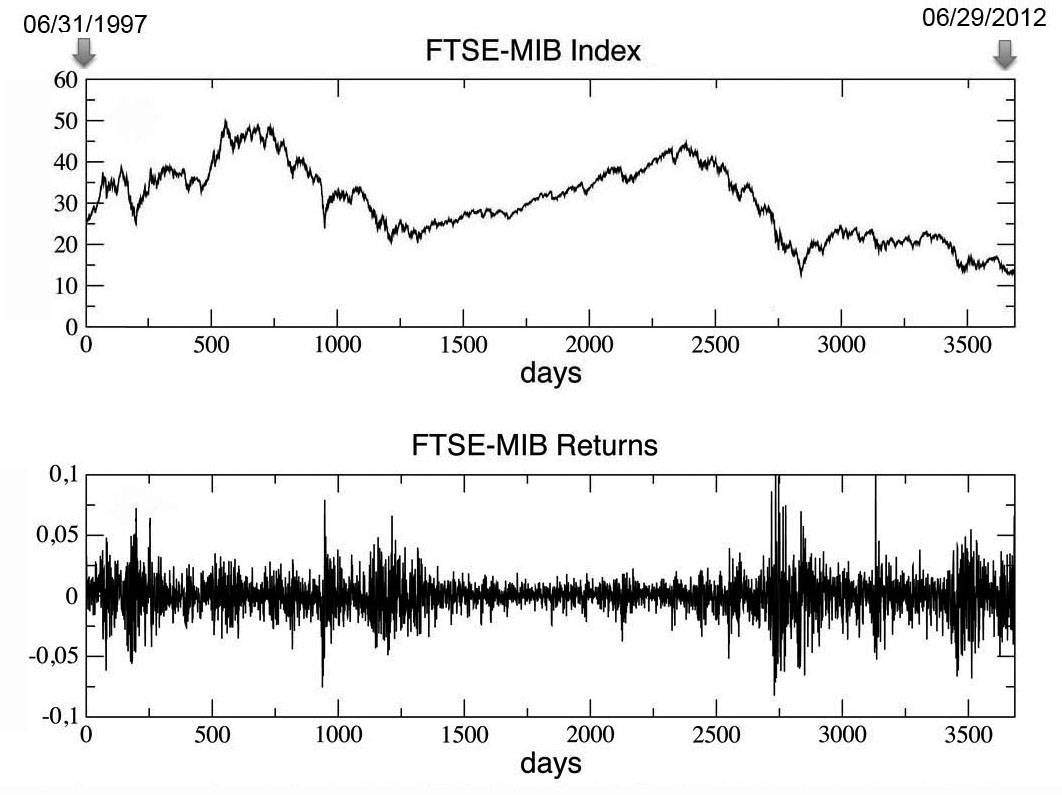,width=10truecm,angle=0}
\epsfig{figure=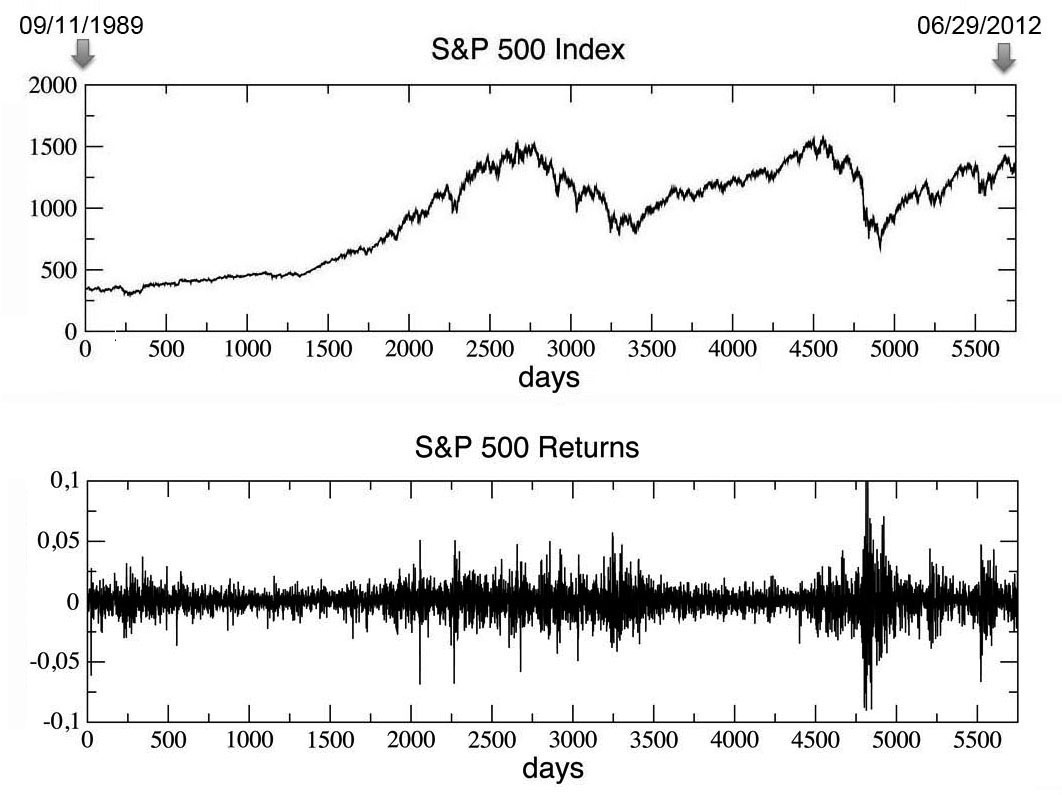,width=10truecm,angle=0}
\end{center}
\caption{{\it Top panels: FTSE MIB All-Share index (from December, 31th 1997 to June, 29th 2012, for a total of $T=3684$ days) and the corresponding series of returns. Bottom panels: S$\&$P 500 index (from September, 11th 1989 to June, 29th 2012, for a total of $T=5750$ days) and the corresponding series of returns. See text for more details.
}}
\label{NEW-time-series}
\end{figure}

\subsection{Details about the Stock Markets considered}

We address two important financial markets: a European one, i.e. the Italian national stock exchange, and the US stock market. In particular, we consider the two following financial indexes:
(a)  the FTSE MIB All-Share index, consisting of the 40 most-traded stock classes on the Milano exchange;
(b) the S$\&$P 500 index, based on the market capitalizations of 500 large companies having common stock listed on the NYSE or NASDAQ.
\\
In Fig.\ref{NEW-time-series} we show the time behavior of the FTSE MIB index, for a period $T$ of $3684$ days, and that of the S$\&$P 500 index, for a period $T$ of $5750$ days, together with their corresponding 'returns' time series. Calling $F_j$ the $j$-th daily value of a given financial index, the returns are defined as the ratio $(F_{j+1}-F_j)/F_j$ (with $j=1,...,T$) and their importance is due to the fact that the standard deviation of the returns in a given time window represents the volatility of the market in that period, i.e. an indicator of what can be  qualitatively called the 'turbulent status' of the market. 
In this respect, the level of volatility clearly influences the possibility of forecasting the market behavior, since it is related to the degree of correlations existing in the financial series \cite{MantegnaStanley,gabaix,preis}.       
\\
An effective way to estimate the presence of correlations in a given time series is the calculation of the time-dependent Hurst exponent  through the so called 'detrending moving average' (DMA) technique \cite{Carbone}.   The DMA algorithm is based on the computation of the following standard deviation $\sigma_{DMA}(n)$ as function of the size $n$ of a time window moving along a financial series $F$ of length $T$: 
\begin{equation}
\sigma_{DMA}(n)=\sqrt{\frac{1}{T-n} \sum_{j=n}^{T} [F_j-\tilde{F}_j(n)]^2},
\label{DMA-equation}
\end{equation}
where  $\tilde{F}_j(n)=\frac{1}{n} \sum_{k=0}^{n-1} F_{j-k}$ is the average calculated in each time window of size $n$, while $n$ is allowed to increase in the interval $[2,T/2]$. 
In general, the function $\sigma_{DMA}(n)$ exhibits a power-law behavior with an exponent $H$ which is precisely the Hurst index of the time series $F$: if $0\le H \le 0.5$, one has a negative correlation or anti-persistent behavior, while if $0.5\le H \le 1$  one has a positive correlation or persistent behavior. The case of $H=0.5$ corresponds to an uncorrelated Brownian process.
\begin{figure}  
\begin{center}
\epsfig{figure=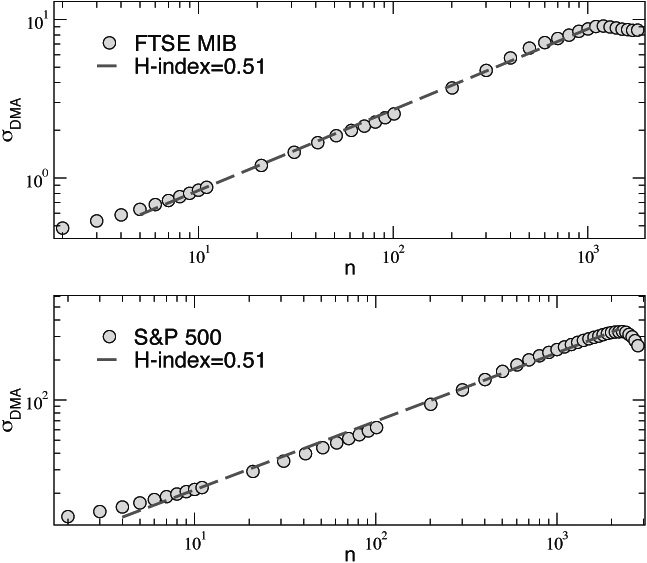,width=7.5truecm,angle=0}
\epsfig{figure=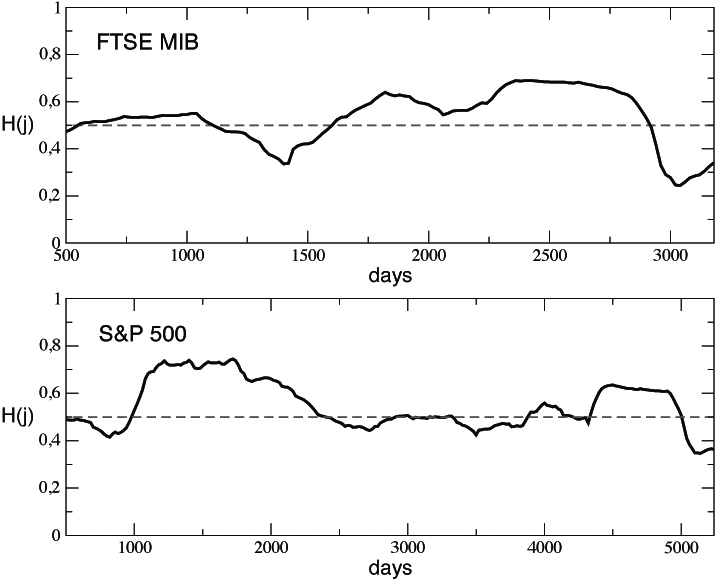,width=8truecm,angle=0}
\end{center}
\caption{ Detrended analysis for the two financial market series shown in Fig.\ref{NEW-time-series}. Left column: the power law behavior of the DMA standard deviation shows an Hurst index oscillating about $0.5$, thus indicating, on average, an absence of correlations over a large time period. Right column: time dependence of the Hurst index for the two series: on smaller time scales, significant correlations emerges. See text for more details.}
\label{DMA}
\end{figure}
\\
In the left column of Fig.\ref{DMA} we show two log-log plots of the $\sigma_{DMA} (n)$, calculated over the complete FTSE MIB and the S\&P 500 time series (gray circles), together with the corresponding power law fits $y \propto n^H$ (dashed lines): in both  cases one observes a Hurst index very close to $0.5$, indicating an absence of correlations on a large time scale. On the other hand, calculating the local value of the Hurst exponent day by day along the time series, significant oscillations around $0.5$ seem to emerge. This is shown in the right column of Fig.\ref{DMA}, where two sequences of Hurst exponent values $H(j)$ (solid lines) are obtained as function of time by considering subsets of the complete FTSE MIB and the S\&P 500 series through sliding windows $W_s$ of size $T_s$, moving along the series with a time step $s$: at each day $j\in[0,T-s]$, we calculate the function $\sigma_{DMA}(n)$ inside the sliding window $W_s$ by substituting $T$ with $T_s$ in Eq.\ref{DMA-equation}, and we compute the corresponding value for $H(j)$ (in Fig.\ref{DMA} we fixed $T_s=1000$ and $s=20$). Taken together, these results seem to suggest that correlations are important only on a local temporal scale, while they cancel out when averaging over long-term periods. As we will see in the following, this feature will affect the performances of the trading strategies considered.   

\subsection{Trading strategies and simulations results}

Of course, the task of our three virtual traders has been very simplified with respect to  reality. Actually, for both the time series considered, they have just to predict, day by day, the upward ('bullish') or downward ('bearish') movement of the index $F_{j+1}$ on a given day with respect to the closing value $F_j$ one day before: if the prediction is correct, we will say that they win, otherwise that they lose. In this regard, we assume that they perfectly know the past history of the indexes, but do not possess any other information and cannot either exert nor receive any influence from the market or from the other traders. At the end of the game, we shall be only interested in comparing the percentage of wins of all the traders, which  of course will depend on the strategy of investment they adopted. 
\begin{figure}  
\begin{center}
\epsfig{figure=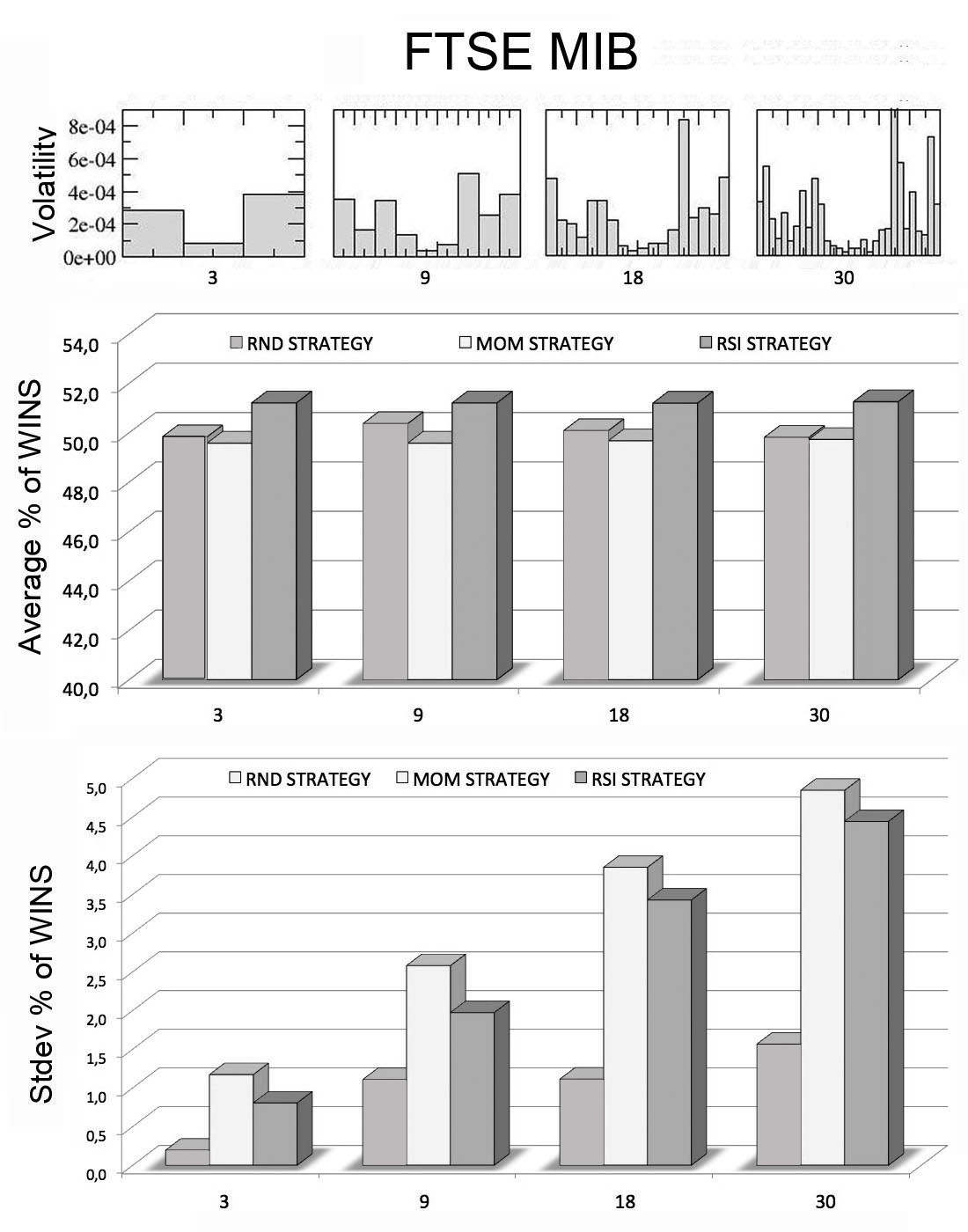,width=9truecm,angle=0}
\end{center}
\caption{Results for the FTSE-MIB All-Share index series, divided into an increasing number of trading-windows of equal size ($3,9,18,30$), simulating different time scales. From top to bottom, we report the volatility of the time series, the percentages of wins for the three strategies over all the windows and the corresponding standard deviations. The last two quantities are averaged over $10$ different runs (events) inside each window. It clearly appears that the random strategy shows, in comparison with the other strategies, a similar average performance in terms of wins, but with smaller fluctuations. See text for further details.}
\label{NEW-Figura-MIB}
\end{figure}
\\
The three possible strategies, each one chosen by a given trader, are the following:

{\it 1) Random (RND) Strategy}
\\
It is the simplest one, since the corresponding trader makes her 'bullish' (index increases) or 'bearish' (index decreases) prediction at the day $j$, for the next day $j+1$, completely at random (just tossing a coin). 

{\it 2) Momentum (MOM) Strategy}
\\
It is a technical strategy based on  the so called 'momentum' $M(j)$, i.e. the difference between the value $F_j$ and the value $F_{j-\Delta j_M}$, where $\Delta j_M$ is a given trading interval (in days). Then, if $M(j)=F_j-F_{j-\Delta j_M}>0$, the trader predicts an increase  of the closing index for the next day (i.e. it predicts that $F_{j+1}-F_j>0$) and vice-versa. In the following simulations we  consider $\Delta j_M=7$ days, since this is one of the most used time lags for the momentum indicator. See Ref. \cite{murphy}.  

{\it 3) Relative Strength Index (RSI) Strategy}
\\
This is also a technical strategy, but it is based on a more complex indicator, called 'RSI', which is a measure of the stock's recent trading strength. Its definition is: $RSI(j)=100-100/[1+RS(j)]$, where $RS(j,\Delta j_{RSI})$ is the ratio between the sum of the positive returns and the sum of the negative returns occurred during the last $\Delta j_{RSI}$ days before $t$. Once  the RSI index, for all the days included in a given time-window of length $T_{RSI}$ immediately preceding the day $j$, has been calculated, the trader who follows the RSI strategy makes her prediction on the basis of a possible reversal of the market trend, revealed by the so called 'divergence' between the original series and the new RSI one. In our simplified model, the presence of such a divergence translates into a change in the prediction of the sign of the difference $F_{j+1}-F_j$, depending on the 'bullish or 'bearish' trend of the previous $T_{RSI}$ days. In the following simulations we choose $\Delta j_{RSI}=T_{RSI}=14$ days, since - again - this value is one of the most commonly used in RSI-based actual trading. See Ref. \cite{murphy}.

\begin{figure}  
\begin{center}
\epsfig{figure=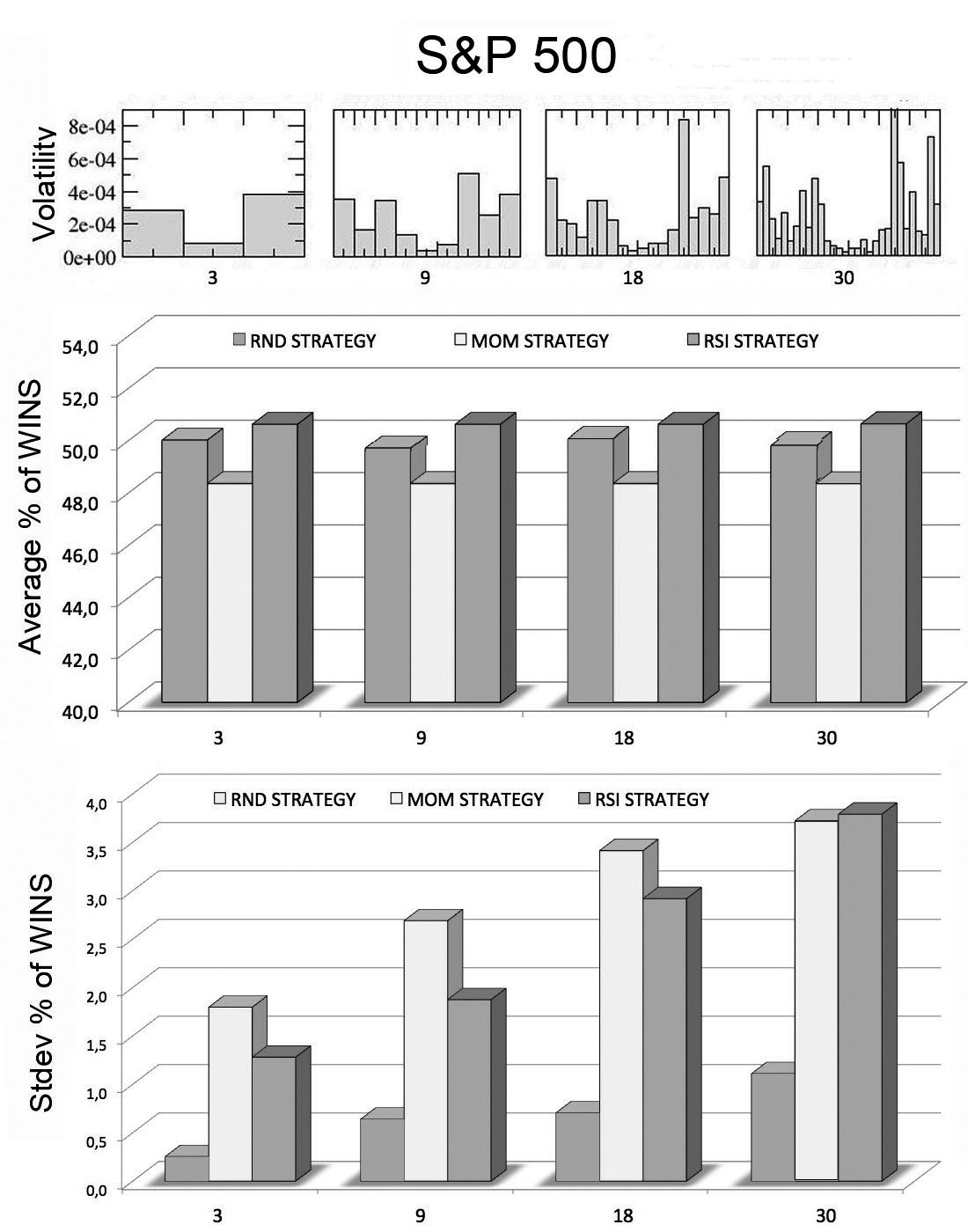,width=9truecm,angle=0}
\end{center}
\caption{Results for the S$\&$P 500 index series, divided into an increasing number of trading-windows of equal size ($3,9,18,30$), simulating different time scales. From top to bottom, we report the volatility of the time series, the percentages of wins for the three strategies over all the windows and the corresponding standard deviations. The last two quantities are averaged over $10$ different runs (events) inside each window. As in Fig.3, the random strategy shows a good average performance and smaller fluctuations.See text for further details.}
\label{NEW-Figura-SP500}
\end{figure}

In order to test the performance of the previous strategies, we divide each one of the two time series (FTSE MIB and S$\&$P 500) into a sequence of $N_w$ trading windows of equal size $T_w=T/N_w$ (in days) and we evaluate the average percentage of wins (with the corresponding standard deviation) of the three traders inside each window while they move along the series day by day, from $j=0$ to $j=T$. This procedure seems advisable since, as we have seen in Fig.\ref{NEW-time-series}, the volatility of both the financial series considered fluctuates significantly  and, as also shown in the previous subsection, the presence of short-term correlations could induce a different behavior of the various strategies at different time scales. 

In Fig.\ref{NEW-Figura-MIB} and Fig.\ref{NEW-Figura-SP500} we show the simulation results obtained for both the FTSE MIB and S$\&$P 500 series. 
For four increasing values of $N_w$, equal to $3$, $9$, $18$ and $30$ windows respectively (corresponding to time periods $T_w$ going from about  $5$ years to about $6$ months), we report (i) the volatility of the returns calculated inside each window, (ii) the average percentages of wins of the three traders calculated over all the trading windows and (iii) the corresponding standard deviations. The quantities (ii) and (iii) are further averaged over $10$ different simulation runs.  
\begin{figure}  
\begin{center}
\epsfig{figure=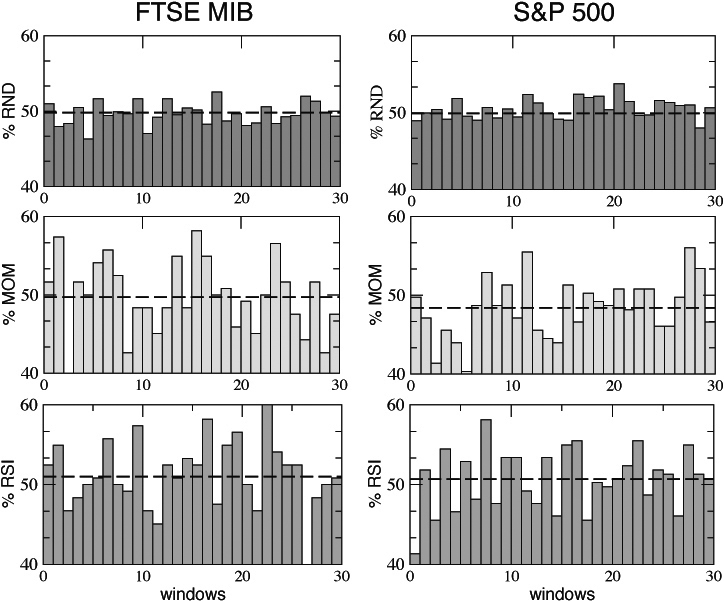,width=10truecm,angle=0}
\end{center}
\caption{The percentage of wins of the different strategies inside each time window - averaged over $10$ different events - is reported, in the case $N_w=30$, for the two markets considered. As visible, the performances of the three strategies (from top to down: RND, MOM and RSI) can be very different one from the others inside a single time window, but averaging over the whole series these differences tend to disappear and one recovers the common $50\%$ outcome shown in the previous figures. See text for further details.
}
\label{NEW-windows-comparison}
\end{figure}
\\
The results for the two time series lead to the same conclusions: on one hand, the long-term average performances of the three strategies in terms of percentage of wins are comparable and restricted in a narrow band just around $50\%$, on the other hand the stability of the random strategy seems always higher than the stability of the technical ones. These global features emerge from the local time behavior of the three traders, which can be better appreciated by plotting their percentage of wins inside each window for the case $N_w=30$, as shown in Fig.\ref{NEW-windows-comparison} for the two time series. Here we see that, on a small time scale, a given strategy may perform much better or worse than the others (probably just by chance, as suggested by Taleb \cite{Taleb}), but the global performances of the three strategies (already presented in the previous figures and here indicated by a dashed line) are very similar and near to $50\%$. In the same figure we can also better see how the fluctuations of the random strategy around the average remain always smaller than those of the other strategies, meaning that, from the point of view of a single trader, the random strategy is less risky than the standard trading ones.
\\
But what would happen if we now imagine to extend the adoption of the random strategy to a larger community of interacting traders investing in a financial market?
\\
This will be addressed in the next section.

\section{Macro effects of random strategies on financial markets stability}

It is well known that financial markets often experience extreme events, like ``bubbles'' and ``crashes'', due to {\it positive feedback} effects which induce sudden drops or rises in prices, in contrast with the {\it negative feedback} mechanism leading to an equilibrium price under normal market circumstances \cite{Buchanan,Helbing-Kern}.
The positive feedback dynamics is strictly related to the presence of avalanches of investments, due to the tendency of human beings to orient themselves following  decisions and behaviors of others (the so called ``herding" effect), particularly in situations where it is not clear what is the right thing to do \cite{HelbingVicsek}. Actually, such conditions are typical for financial markets, in particular during volatile periods. Remarkably, in this context, bubbles and crashes may reach any size and  their probability distribution follows a power law behavior 
\cite{mandelbrot2,MantegnaStanley,PeinkeNature,Farmer1,Bouchaud2,Sornette,Helbing-Nature,krawiecki,parisi}.  
\\
In this section we show that, assuming information cascades between agents \cite{Bikhchandani-Hirshleifer-Welch} as the underlying mechanism of financial avalanches, it is possible to obtain a power law distribution of bubbles and crashes through a self-organizing criticality (SOC) model implemented on a given network of technical traders investing in a financial market. Moreover, we also show that it is possible to considerably reduce the maximum size of these avalanches by introducing a certain percentage of traders who adopt a random investment strategy.   
\\
Our model is inspired by the SOC phenomenon observed in many physical, biological and social systems \cite{Bak}, and, in particular, in  the Olami-Feder-Christensen (OFC) model \cite{Olami,Caruso} that has been proposed to study earthquakes dynamics \cite{Sornette,Mantegna1}. In our implementation, we identify the OFC earthquakes with the herding avalanches of investments observed in financial markets, therefore we called our model the ``Financial Quakes" (FQ) model. 

\begin{figure}  
\begin{center}
\epsfig{figure=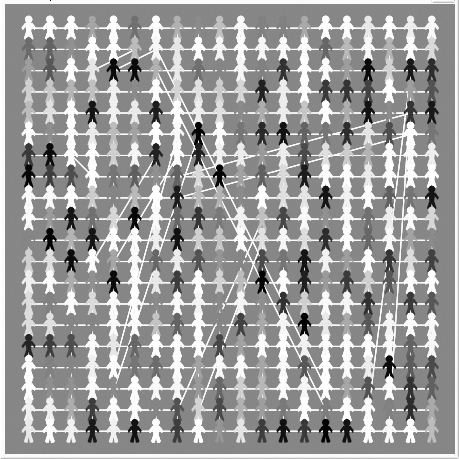,width=10truecm,angle=0}
\end{center}
\caption{An example of small-world $2D$ network with $N=400$ traders, smaller but similar to that one used in our numerical simulations (where we consider $N=1600$ agents). Different colors indicate the different values of information possessed by the various agents at the beginning of the simulation. See text for further details.}
\label{SMALL-WORLD}
\end{figure}

\subsection{FQ model on a Small-World 2D lattice} 

Let us consider a small-world (SW) undirected network of interacting traders (agents) $A_i$ ($i=1,2,...,N$), obtained from a regular $2D$ lattice (with open boundary conditions) by means of a rewiring procedure (with a rewiring probability $p=0.02$) which transforms short range links into long range ones \cite{Caruso3}, thus preserving the clustering properties of the network and its average degree, i.e. the average number of nearest neighbors of each node (see Fig.\ref{SMALL-WORLD}). 
In the following we consider a total of $N=1600$ agents, with an average degree $<k>=4$.
As in the previous section, each agent may invest in a given market by using a given trading strategy. Again, the traders may bet their money trying to guess the bullish or bearish daily behavior of the FTSE MIB or of the S$\&$P 500 indexes. However, the presence of mutual interactions does impose several updates to the old investment mechanism, which now has to be able to take into account the possibility of herding effects.
\\
For this purpose, we have imagined that, at each simulation time $t$, all the agents have a certain quantity of information $I_i(t)$ about the market considered. Initially, at $t=0$, it assumes a random value in the interval $(0,I_{th})$, where $I_{th}=1.0$ is an arbitrary threshold equal for all the traders. When the simulation starts, i.e. for $t>0$, information may change due to the following two mechanisms. 
\\
\begin{figure}  
\begin{center}
\epsfig{figure=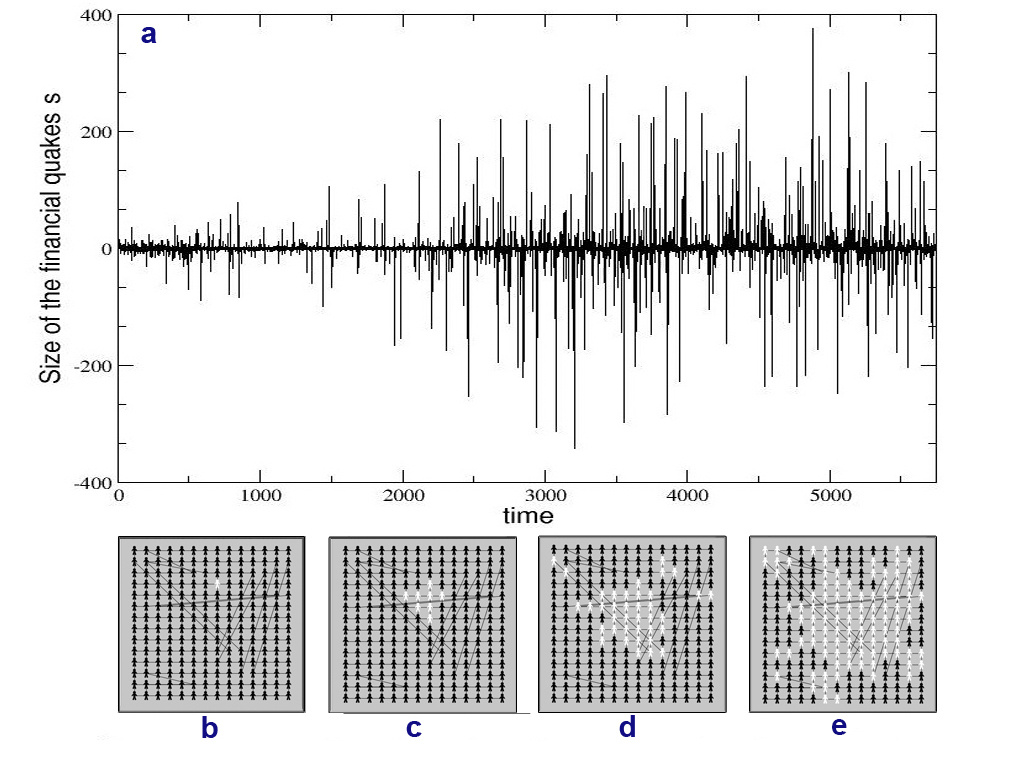,width=16.5truecm,angle=0}
\end{center}
\caption{(a) An example of time series  of ``financial quakes'' for a SW network of traders investing in a given stock market (bottom panel). Both positive cascades (``bubbles'') and negative avalanches (``crashes'') are visible. (b-e) Time evolution of a single financial quake in the small-world network: starting from a single active trader (b), the herding-activated avalanche (in white) rapidly reaches different part of the network (c-d-e), helped by the presence of long range connections. See text for further details.}
\label{NEW-financial-quakes}
\end{figure}
- The first one is a {\it global one}: due to external public  information sources, all the variables $I_i(t)$ are simultaneously increased by a quantity $\delta I_i$, different for every agent and randomly extracted within the interval $[0,(I_{th}-I_{max}(t))]$, where $I_{max}(t)= \max \{I_i(t)\}$ is the maximum value of the agents' information at time $t$. If, at a given time step $t^*$, the information $I_k(t^*)$ of one or more agents $\{A_k\}_{k=1,...,K}$ exceeds the threshold value $I_{th}$, these agents become ``active" and take the decision of investing, i.e. they bet on the behavior of a given financial index value $F_j$ compared to that one of the day before $F_{j-1}$.
\\
- The second one is a {\it local} one and depends on the topology of the network: as  they invest on the market, all the active traders $\{A_k\}_{k=1,...,K}$ will also share their information with their neighbors according the following herding mechanism inspired by the stress propagation in the OFC model for earthquakes:
\begin{equation}   
\label{av_dyn}       
I_k > I_{th}  \Rightarrow \left\{ 
	\begin{array}{l}
       I_k \rightarrow 0, \\
       I_{nn} \rightarrow I_{nn} + \frac{\alpha}{N_{nn}} I_k        \end{array} 
	\right.
\end{equation}
where ``nn'' denotes the set of nearest-neighbors of the agent $A_k$ and $N_{nn}$ is the number of her direct neighbors. Of course, the neighbors that, after receiving this {\it surplus} of local information, exceed their threshold, become active too and will invest imitating the investment of agent $A_k$. In turn, they will also transfer their information to their neighbors thus activating a positive feedback process which could be able to generate an avalanche of identical investments, i.e. what we call a financial quake.
\\
The parameter $\alpha$ in Eq.\ref{av_dyn} controls the level of dissipation of the information during the dynamics ($\alpha=1$  means no dissipation) and it is fundamental in order to drive the system in a SOC-like critical state. In analogy with the OFC model on a SW network \cite{Caruso,Caruso3} we set here $\alpha=0.84$, i.e. we consider some loss of information during the herding process. This value ensures the emergence of avalanches that can reach any size $s$, as  shown in Fig.\ref{NEW-financial-quakes}. Let us explain in detail what we plot in this figure.
   
\par\begin{figure}
\begin{center}
\epsfig{figure=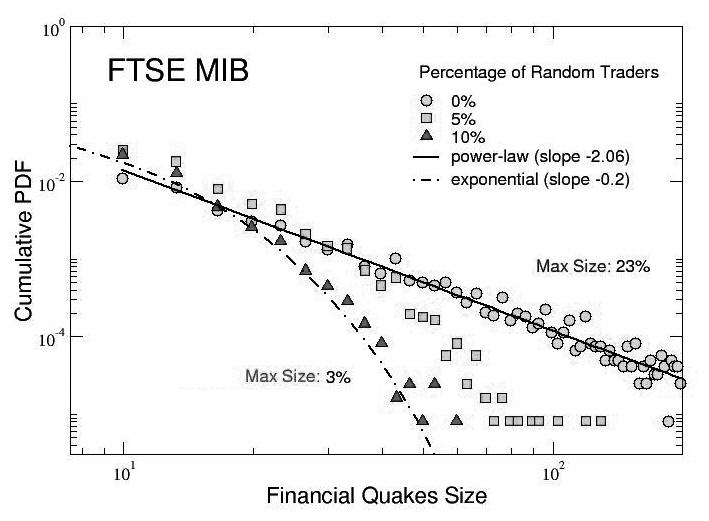,width=10.5truecm,angle=0}
\epsfig{figure=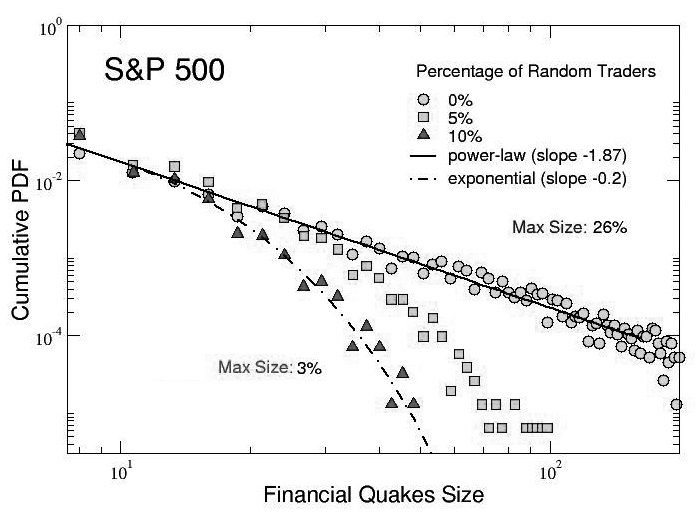,width=10.5truecm,angle=0}
\end{center}
\caption{
Distributions of the absolute values of the size of herding avalanches occurring in the small-world (SW) community of investors, with and without random traders, for both the FTSE MIB index (top panel) and the S$\&$P 500 index (bottom panel). Each curve has been cumulated over 10 different events. In the absence of random traders, i.e. with only RSI traders (circles), the distributions follow a well defined power law behavior. On the other hand, increasing the amount of random traders, in particular with percentages of $5\%$ (squares) and $10\%$ (triangles), the distributions tend to become exponential. See text for further details.}
\label{NEW-financial-quakes-pdf} 
\end{figure}
In the bottom panels of the figure we show, from left to right (b-e), a sequence of four snapshots reproducing the time evolution of a financial quake.  
In the snapshot (b) we see that, at a given simulation time $t^*$, a certain agent $A_k$, situated in the right part of the network and colored in white, overcomes her information threshold and gives rise to the avalanche.
In this example we consider only technical traders, i.e. all the agents adopt the RSI strategy introduced in the previous section (this also means that, if $K>1$, i.e. if more than one trader initially overcomes the threshold, all of them will make the same prediction about the market). Following this strategy, the agent $A_k$ will make her prediction $P_j$ (positive or negative) about the sign of the difference $(F_j-F_{j-1})$ of a given financial index at day $j$ (note that the daily index $j$ of the financial series does not coincide, here, with the simulation time $t$ but it is updated at the beginning of every financial quake).  
Then, following the herding rule (\ref{av_dyn}), the same agent will transfer her information to her neighbors, some of which will, in turn, overcome their threshold and will invest simultaneously adopting the same prediction $P_j$ of the first agent, as visible in snapshot (b). This process goes on iteratively (see snapshots (c) and (d)) until there are no more active agents in the system (i.e. when $I_i < I_{th}$ $\forall i$). Then, the financial quake is over and the prediction $P_j$ is finally compared with the sign of $(F_j - F_{j-1})$ in the time series: if they are in agreement, all the agents who have contributed to the avalanche win, otherwise they lose. In the former case the size $s_j$ of the quake $j$-th (i.e. the total number of agents involved in that quake) will have a positive sign (bubble), in the latter a negative one (crash). The time sequence $\{s_j\}_{j=1,...,T}$ of these financial quake sizes during a single simulation run is plotted in the top panel (a) of Fig.\ref{NEW-financial-quakes}: it is evident that, after a short transient, the system rapidly reaches a critical-like state where bubbles and crashes of any size are observed. In fact, the log-log distribution of the absolute value of the avalanches' size results to be a power law for both the FTSE MIB and the S$\&$P 500 time series, as shown in the next figure. 
\\ 
In Fig.\ref{NEW-financial-quakes-pdf} we report the probability distribution $P_N (s)$ of the absolute value of the size of the financial quakes, cumulated over 10 events, for the FTSE MIB (top panel) and S$\&$P 500 (bottom  panel) financial series. When only RSI traders are present, as in the case we are addressing now, a well defined power law behavior is observed for both the series (circles), as demonstrated by the corresponding fits (solid lines) with exponents, respectively, $-2.06$ and $-1.87$. In the same figure we also report how these distributions are affected by the introduction, in the SW network, of an increasing percentage $P_{RND}$ of random traders, uniformly distributed among the $N=1600$ RSI agents. However, this case needs further clarifications. 
\\
The main point to take into account in order to understand the effect of random trading in a community of technical investors is that, in contrast to RSI ones,  random traders (i) are not activated by their neighbors, precisely because they invest at random, and (ii) they do not activate their neighbors, since a random trader has no specific information to transfer. In other words, random traders only receive the information $\delta I_i$ from external sources, but do not exchange any information with other agents. In terms of dynamics, we simply set $\alpha = 0$ for random traders in Eq.\ref{av_dyn}. This means that, even if random traders can invest exactly in the same way as the other agents when they overcome their information threshold,  they do not take part in the herding process, therefore they are not involved in any financial quake.
\\ 
Remarkably, as visible in both the panels of Fig.\ref{NEW-financial-quakes-pdf}, the consequences of the introduction of even a small percentage ($5\%$ or $10\%$) of these random traders are that the size of the financial quakes is immediately reduced and the corresponding probability distributions tend to become exponential, as also confirmed by the fits (dot-dashed lines). In particular, in both the cases the maximum size of the herding avalanches reduces from about $25\%$ of the entire network, obtained with RSI traders only, to less than $3\%$ , with $10\%$ of random traders.    

\begin{figure}  
\begin{center}
\epsfig{figure=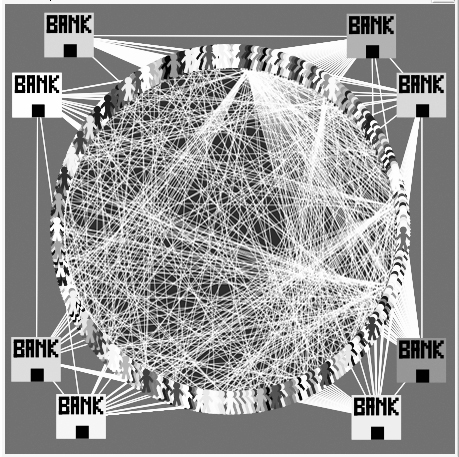,width=10truecm,angle=0}
\end{center}
\caption{An example of scale-free network with $N=400$ nodes, the great majority of whom are RSI traders and only $8$ are random traders (here indicated as Banks), coinciding with the main hubs of the network (i.e. nodes with more than a given number of links - in this example more than $25$). In our simulations we adopt a similar network but with $N=1600$ nodes. See text for further details.}
\label{SCALE-FREE}
\end{figure}

\subsection{FQ model on a Scale-Free network} 

We have seen, therefore, that a relatively small number of random investors distributed uniformly at random in our network  is able to suppress dangerous herding-related avalanches, thus leading the system out of the critical-like state.
This happens in the context of a trading community where agents are distributed over a regular $2D$ lattice with small-world topology. 
But, in such a network, all the agents are equivalent, i.e. they all have, more or less, the same number of neighbors (four, on average). 
It is interesting to investigate what would happen if the percentage of random traders is decreased  further, but if, at the same time, their importance, in terms of connectivity within the network, is increased.
\\
In this section we try to answer this question, by adopting another kind of network with a different topology. In particular, we choose an undirected scale-free (SF) network, i.e. 
an example of a network displaying a power-law distribution $p(k)\sim
k^{-\gamma}$ in the node degree $k$. By using the preferential attachment growing procedure introduced in \cite{barabasi}, we start from $m+1$ all to all connected nodes and at each time step we add a new node with $m$ links. These $m$ links point to old nodes with probability $p_i=\frac{k_i}{\sum_j k_j}$, where $k_i$ is the degree of the
node $i$. This procedure allows a selection of the $\gamma$ exponent of the power law scaling in the degree distribution with $\gamma=3$ in the thermodynamic limit ($N \longrightarrow \infty$).
\\
In Fig.\ref{SCALE-FREE} we report a useful visualization of a SF network, where all the nodes (traders) are put on a circle except the {\it hubs}, i.e. the mostly connected nodes, which are put out of the circle and, in a financial context, could represent banks or great investors. In our simulations we adopt a SF network with $N=1600$ traders and we consider two possibilities: (i) only technical RSI traders and (ii) a majority of technical RSI traders plus a small number $N_H$ of random traders represented by the main hubs of the network (e.g. all the nodes with $k>50$). The question, now, is: will these $N_H$ hyper-connected random traders be able, alone, to reduce the size of financial quakes?
\\
\begin{figure}
\begin{center}
\epsfig{figure=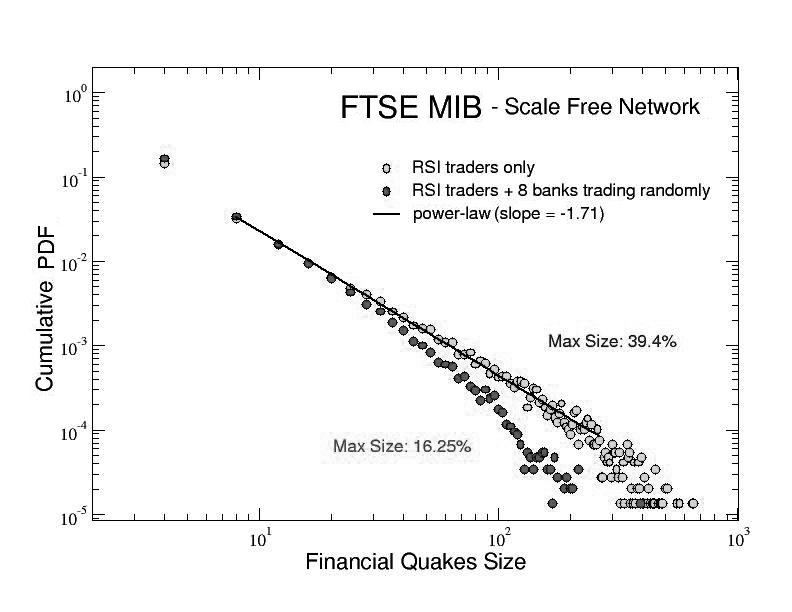,width=10.5truecm,angle=0}
\epsfig{figure=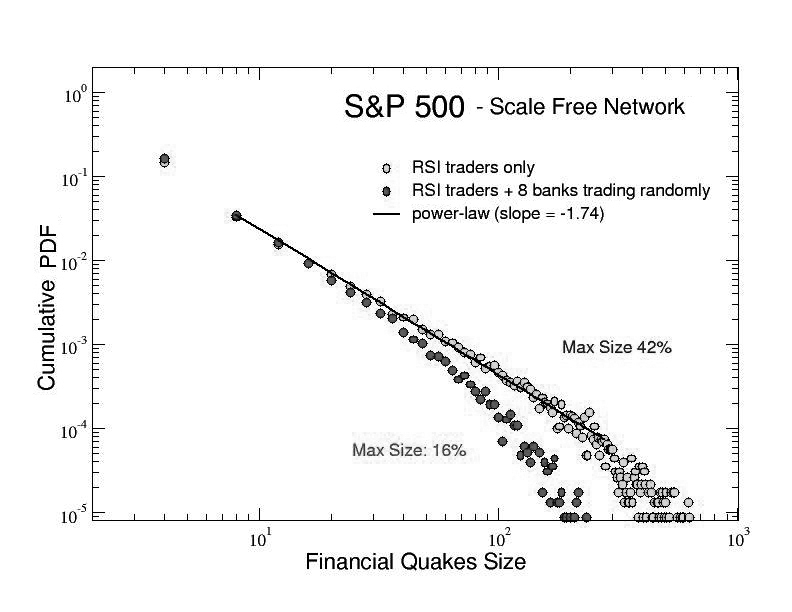,width=10.5truecm,angle=0}
\end{center}
\caption{
Distributions of the absolute values of the size of herding avalanches occurring in the scale-free (SF) community of $N=1600$ investors, with and without random traders, for both the FTSE MIB index (top panel) and the S$\&$P 500 index (bottom panel). Each curve has been cumulated over 10 different runs. In the absence of random traders, i.e. with RSI traders only (gray circles), the distributions follow a well defined power law behavior. The introduction of just a few number of random traders (about $8$ over $1600$), corresponding with the hyper-connected nodes of the network (i.e. with $k>50$), it is enough to dampen the avalanches reducing their maximum size to less than $40\%$ (black circles). See text for further details.}
\label{NEW-scale-free-financialquakes} 
\end{figure}
First of all, we should check if the dynamics of our financial quakes model on a scale-free network with only RSI traders is still able to reach a SOC-like critical state, as for the small world $2D$ lattice. We find that this system does show power law distributed avalanches of investments, providing that the parameter $\alpha$ in Eq.\ref{av_dyn} is slightly increased with respect to the value adopted in the previous section. In particular, we use here $\alpha=0.95$. Then, we perform two sets of simulations, with $10$ different runs each, either in absence or in presence of the $N_H$ hyper-connected random traders. 
\\
In Fig.\ref{NEW-scale-free-financialquakes} we show the corresponding results cumulated over $10$ runs for both the FTSE MIB (top panel) and the S$\&$P 500 (lower  panel) series, analogous to those presented in Fig.\ref{NEW-financial-quakes-pdf}. 
Power law distributions with exponents $-1.71 $ and $-1.74 $ emerge respectively for the two time series considered, when only  RSI traders are considered, indicating a SOC-like behavior as seen in the previous section for a SW topology. 
It is evident that, in both the cases, the presence of just (on average) $N_H=8$ hubs investing at random is enough to alter the power law distributions, damping the herding-related positive feedbacks and reducing the avalanches' size to less than $40\%$ of their original value (we check that such a reduction disappears if the random traders are no more the hubs of the SF network but $8$ randomly chosen nodes). 
More precisely, comparing these results with the analogous ones presented for the small-world network, we find that here, with about $8$  hyper-connected random traders, we yield the same damping effect obtained in the previous case with  $2\%$ of equivalent (in terms of connectivity) random traders uniformly distributed on the $2D$ lattice. Since $2\%$ of $1600$ corresponds to $32$ traders, we can conclude that,
compared to a small-world configuration of the network, with the scale-free topology a fewer number of random investors (provided that they are the mostly connected agents) are needed in order to dampen bubbles and crushes. This evidence could suggest some policy implication, later addressed in the conclusions.

\subsection{Capital gains and losses in the FQ model} 

It is particularly interesting, at this point, to focus on the personal gains or losses that the interacting agents experience during the whole trading period considered. 
\\
In the simulations presented in section 4, the three traders always  invested in the market the same amount of a virtual capital, i.e. one credit unit, starting from an initially null wealth  (in the following we use the term "capital" and "wealth" as synonymous). Therefore, the deviation of their final average percentage of wins from $50\%$ could represent by itself a measure of their final, positive or negative, capital. In that section we showed that, from the micro perspective of a single, isolated investor, the adoption of a random trading strategy could be as much profitable as the technical ones but, at the same time, also much less risky. We see, now, how the interaction among traders, realized through different network topologies, affects those results from the point of view of the wealth distribution.               
\\
In order to make the mechanism of investment more realistic, we refine the FQ dynamics in the following way. We assign at the beginning of each simulation exactly the same initial capital $C$ of $1000$ credits to all the traders, then we let them invest in the market (when established by the dynamical rules) according to the following prescriptions:  

- the first bet of each agent does not modify her capital;

- if an agent wins thanks to a given bet (for example after being involved in a given, big or small, positive financial quake), in the next investment he will bet a quantity $\delta C$ of money equal to one half of her total capital $C$ at that moment, i.e.  $\delta C=0.5C$;

- if an agent loses due to an unsuccessful investment (for example after a negative financial quake), the next time he will invest only ten percent of her total capital at that moment, i.e. $\delta C=0.1C$. 

\begin{figure}
\begin{center}
\epsfig{figure=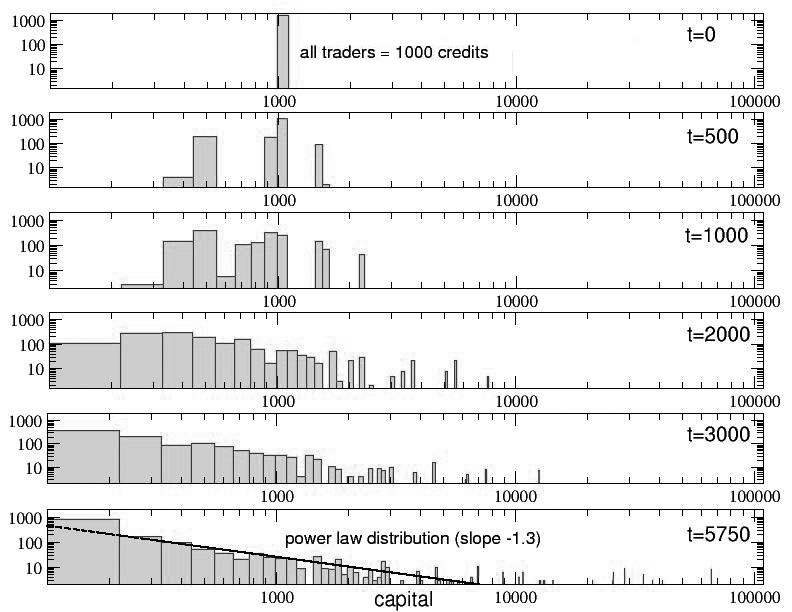,width=10truecm,angle=0}
\end{center}
\caption{Time evolution (from top to bottom) of the distribution of the individual capital of a small-world $2D$ lattice of RSI traders investing on the S$\&$P 500 stock market. Starting from the same initial capital of $1000$ credits each, most of the traders quickly lose money, while a small minority of them considerably increases its capital until, at the end of the simulation, the global wealth distribution becomes a Pareto-like power law. See text for further details.        
}
\label{NEW-capital-time-evolution-RSI-only} 
\end{figure}
\begin{figure}
\begin{center}
\epsfig{figure=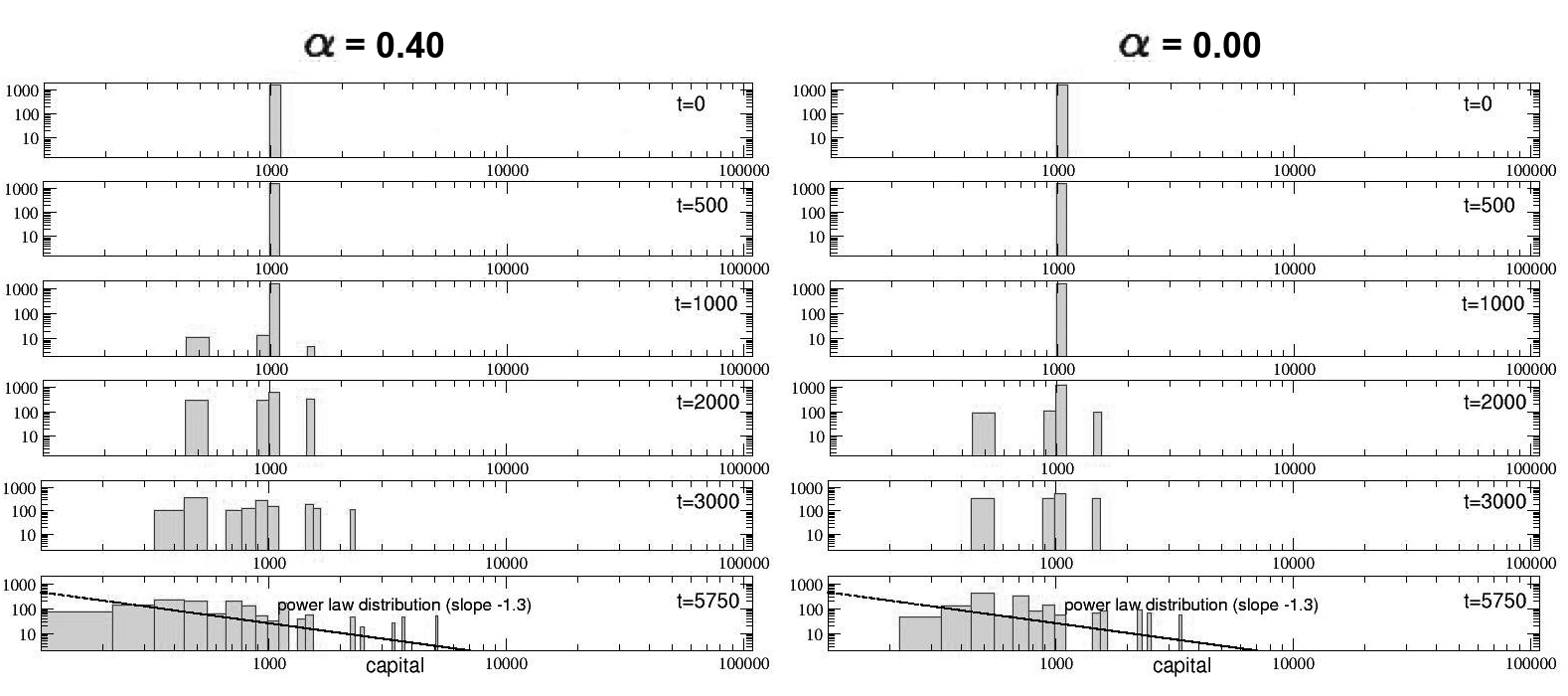,width=16.3truecm,angle=0}
\end{center}
\caption{\label{NEW-asymmetric-capital-distribution} 
The same time evolution as in the previous figure, but for two decreasing values of the parameter $\alpha$, which controls the dissipation of the informative flow in the FQ dinamycs. In particular, we set $\alpha = 0.40$ (left panel) and $\alpha = 0.00$ (right panel). As visible, in both cases the Pareto-like power law distribution of capital observed for $\alpha = 0.84$ disappears. See text for further details.
}
\label{NEW-asymmetric-capital-distribution-transf} 
\end{figure}

Due to these rules, after every financial quake, the capital of all the active agents involved in the herding-related avalanche will increase or decrease by the quantity $\delta C$. On the other hand,  
the wealth of random traders, who do not take part in avalanches, can change only when they overcome their information threshold due to the external information sources.
\\
Let us consider, first, the small-world $2D$ lattice topology. In Fig.\ref{NEW-capital-time-evolution-RSI-only} we show a single-event time evolution of the capital distribution $P(C)$ of $N=1600$ RSI traders, investing in the S$\&$P 500 market. It is clearly visible that, even if they start all with the same initial capital of $1000$ credits, many of them quickly lose most of their money while, on the other hand, a small group of lucky agents largely increases the capital until, at the end of the simulation, the resulting capital distribution is a Pareto-like power law, with exponent $-1.3$. We verified that this distribution is very sensitive to the herding mechanism among RSI traders: in fact, reducing the informative flow among them, i.e. decreasing the value of $\alpha$ in Eq.\ref{av_dyn} from $0.84$ to $0.40$ and then to $0.00$, the system tends to exit from the critical-like state, avalanches are drastically reduced and final capital inequalities disappear (as visible in Fig.\ref{NEW-asymmetric-capital-distribution-transf}).
\begin{figure}
\begin{center}
\epsfig{figure=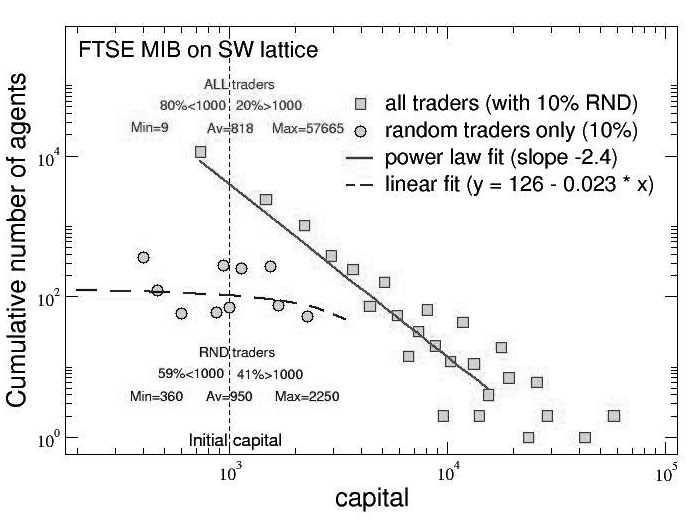,width=10.5truecm,angle=0}
\epsfig{figure=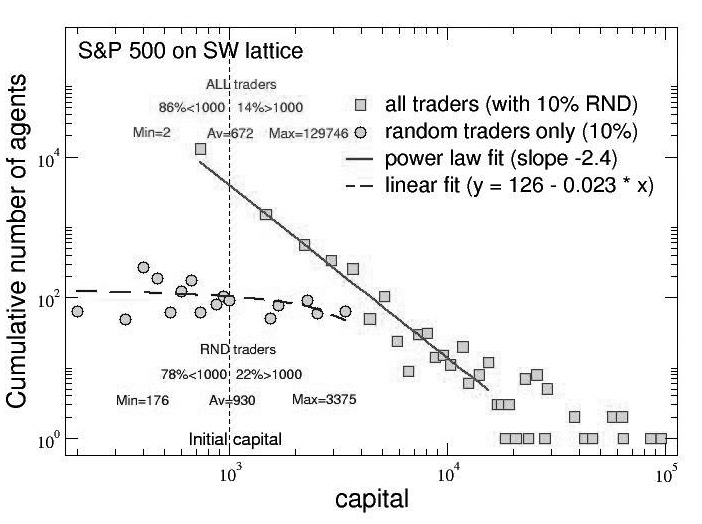,width=10.5truecm,angle=0}
\end{center}
\caption{\label{NEW-asymmetric-capital-distribution} 
Capital/wealth distribution for a small-world $2D$ lattice with a $10\%$ of random traders over a total of $N=1600$ RSI traders, calculated at the end of the simulation and cumulated over 10 different realizations. Both the FTSE MIB (top panel) and the S$\&$P 500 (bottom panel) time series are considered. While the global capital distribution of all the traders is, in both  cases, a Pareto-like power law with exponent $-2.4$ (squares), the capital distribution of the random traders only (circles), though fluctuating, stays almost constant. Details of the distributions are also reported
in the figure. See text for further details.            
}
\label{NEW-asymmetric-capital-distribution-SW} 
\end{figure}
\\
Cumulating data over 10 different simulations, with the same initial condition about capital, and introducing a $10\%$ of random traders among the RSI ones, we focus now on the final capital distribution and obtain the results presented in the two panels of  Fig.\ref{NEW-asymmetric-capital-distribution-SW}, for both the FTSE MIB and the S$\&$P 500 time series.
\\
In the top panel, corresponding to the FTSE MIB series, the final global capital distribution shows again a Pareto-like power law behavior, with exponent $-2.4$, but the partial distribution of the random traders only is completely different, staying almost constant (apart from the fluctuations) or, more precisely, decreasing linearly with an angular coefficient equal to $0.023$ \cite{nota}. 
Looking at  the details of the simulation, we discover something interesting: while, globally,  $80\%$ of traders have, in the end, a lower capital with respect to the initial one of $1000$ credits, the same holds for only $59\%$ of random traders. Moreover, the average final capital of all the traders is of $818$ credits, against a higher average capital of random traders only, equal to $950$ credits. This means that, on average, random trading seems more profitable with respect to the RSI one. On the other hand, the final range of capital is very different in the two cases: for all the traders, the final capital goes from a minimum of $9$ credits to a maximum of $57665$, while, for random traders only, it goes from $360$ to $2250$ credits. In other words, as expected, the random trading strategy seems also much less risky than the technical one: actually, while it is true that the RSI strategy allows a very large gain for a very small number of lucky traders (only $0.6\%$ of the total ends with more than $10000$ credits), it also yields substantial losses for the majority of agents; on the contrary, while the best random traders can gain much less than the best RSI ones, the less lucky of them can also lose little (on average, about $48\%$ of the RSI traders lose more than the worst random one).  
\begin{figure}
\begin{center}
\epsfig{figure=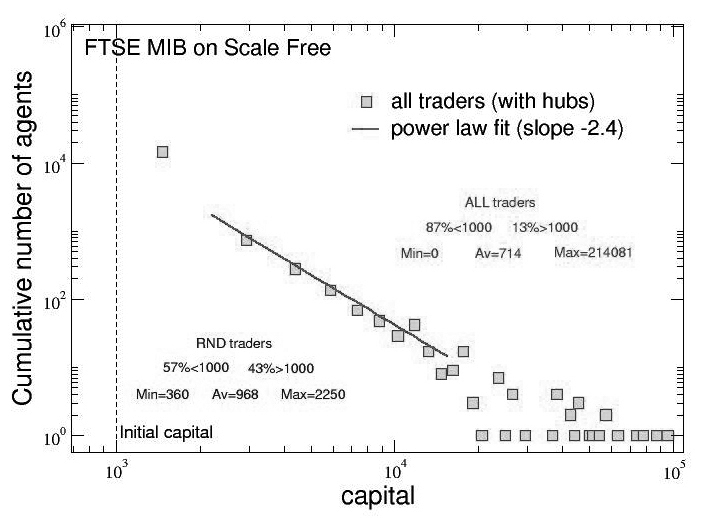,width=10.5truecm,angle=0}
\epsfig{figure=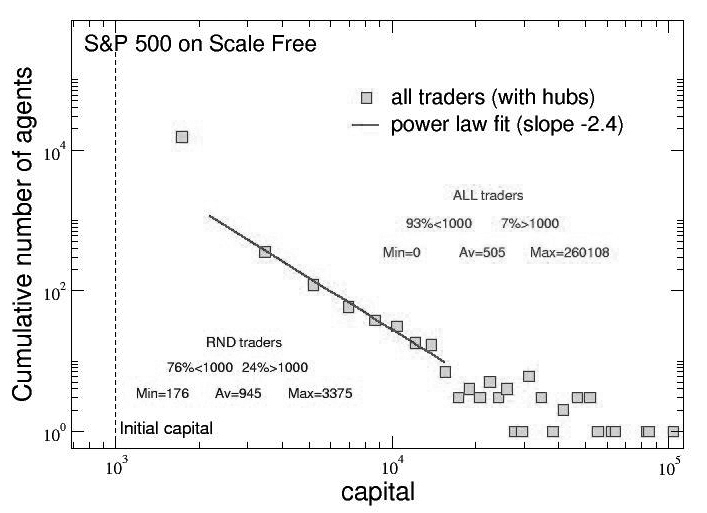,width=10.5truecm,angle=0}
\end{center}
\caption{\label{NEW-asymmetric-capital-distribution} 
Capital/wealth distribution for a scale-free network with, in average, $8$ hyper-connected random traders over a total of $N=1600$ RSI traders, calculated at the end of the simulation and cumulated over 10 different realizations. Both the FTSE MIB (top panel) and the S$\&$P 500 (bottom panel) time series are considered. The global capital distribution of all the traders is, in both the cases, a Pareto-like power law with exponent $-2.4$ (squares). The distribution of the random traders does not appear given their small number. Details of the distributions are also reported in the figure. See text for further details.    
}
\label{NEW-asymmetric-capital-distribution-SF} 
\end{figure}
\\
Similar results are shown in the bottom panel for the S$\&$P 500 time series. A Pareto-like power law with exponent $-2.4$ has been found, again, for the final capital distribution of all the traders, while a fluctuating linear behavior (with the same very small negative slope as before) characterizes the capital distribution of random traders only. Details on the distributions are the following: $86\%$ of all the traders have, at the end of the simulations, less than the initial capital, with an average capital of $672$ credits, a minimum of $2$ and a maximum of $129746$; for the random traders only, $78\%$ of them have at the end less than $1000$ credits, with an average capital of $930$, a minimum of $176$ and a maximum of $3375$. Again, random trading seems to be much less risky and, on average, more profitable than the technical strategies, even if a small number of very lucky RSI agents can become really very rich (in this case more than before, even if, again, only about $0.6\%$ of the traders ends with more than $10000$ credits; on the other hand, about $46\%$ of RSI traders lose more than the worst random one).                      
\\
Finally, let us consider the scale-free network topology, introduced in the previous section. The initial conditions for the capital distribution are  always the same, $1000$ credits for all the agents, but now we have only a very small number $N_H$ of hyper-connected random traders (hubs) distributed among the RSI ones. In our SF networks with $N=1600$ nodes, the average number of hubs with more than $50$ neighbors is $8$, therefore we have on average  $0.5\%$ of random traders, but representing very important investors, who occupy a privileged position in the trading community.
In this case, the results for the final wealth distribution are shown in Fig.\ref{NEW-asymmetric-capital-distribution-SF} and are analogous to those shown in the previous figure for the SW lattice. 
In fact, the final capital distribution of all the traders, cumulated for 10 runs, is again a power law with exponent $-2.4$ for both the FTSE MIB (top panel) and the S$\&$P 500 (bottom panel) time series. The final distribution for random traders only is not plotted since it is statistically not significant due to their small number, but we report details for both the distributions. 
\\
Actually, we see that, for the FTSE MIB series, the $87\%$ of all the traders have, at the end, less than their initial capital, with an average of $714$ credits, a minimum of $0$ and a maximum of $214081$. On the other hand, only $57\%$ of the random traders have lost more than they gained, with an average of $968$ credits, a minimum of $360$ and a maximum of $2250$. Similarly, for the S$\&$P 500,  $93\%$ of all the traders end with less than they had at the beginning, with an average capital of $505$ credits, a minimum of $0$ and a maximum of $260108$, while, among the random traders, $76\%$ lose, with an average of $945$ credits, a minimum of $176$ and a maximum of $3375$. 
\\
\indent
Summarizing, also for the scale-free network topology we find that the random strategy is, at the end of the game, more profitable and less risky. In this case, a very small number of hubs trading at random results not only able to reduce financial avalanches (bubbles and crashes), but can also do this without great risks in terms of capital.     
On the other hand, if a small number of technical traders can gain very much (only about $ 0.9\%$ of all the RSI traders end with more than $10000$ credits for both the FTSE MIB and the S$\&$P 500 series), the large majority of them,  about  $70\%$, lose more than the worst random traders. 
\\
Comparing these results with the corresponding ones found for the small-world topology, where all the traders had, on average, the same connectivity, we can conclude that the scale-free topology, with an unequal (power law distributed) connectivity, amplifies  the inequalities in the final distribution of wealth: the richest traders in the SF network double the richest ones in the SW network, while only in the SF network we find a not negligible percentage of particularly unlucky traders - $4\%$ for the FTSE MIB and even the $10\%$ for the S$\&$P 500 - that lose {\it all} their initial capital.

\section{Conclusive remarks and Policy Suggestions}

We have reviewed some results from recent investigations about the positive role of randomness in socio-economic systems. The greatest attention has been given to the description of consequences deriving from the adoption of random strategies in financial markets. 
From the microeconomic perspective, it has been shown that, if investors chose a completely random approach to decide about their investments, instead of costly and arbitrary technical strategies, they would end up, in average, with the same wealth, but they would incur in  a much lower risk. 
Further, from a macroeconomic point of view, we presented several  results which show   how  that the only  existence of  a few random investors in a 2D small-world lattice,  representing a trading community of interacting agents, reduces substantially the magnitude of financial avalanches. 
\\
After this, we pushed forward our analysis by showing how  a different network structure does not change our general findings and, more relevantly, how  just a very small number of hyper-connected investors (i.e. hubs in a scale-free network of traders) are required in order to obtain a stabilizing effect and dampen financial bubbles and crashes. This result is particularly encouraging, since it reinforces our suggestion for a very specific policy: financial market stabilization is possible if participants are told and convinced that they should not rely completely on the signals they receive from other investors. 
If, for example, under the surveillance of a central banker, a very small number of important investment banks chose to make their investments randomly, other traders would immediately stop interpreting signals and would immediately give up in their continuous seek for confirmation and credibility. Prophecies would no longer be legitimate and followed, since everybody would know that the possibility of random investments do exist. This would immediately stabilize the market and fade out speculations. 
It remains to be investigated  the dynamics of  artificial markets with feedback mechanisms, a natural step towards a self-regulating, participatory market society \cite{Economics2.0}.

\section{Acknowledgments}

We would like to thank Dirk Helbing for many fruitful discussions and for his contributions to the research project to which this paper belongs.



\begin{thebibliography}{10}


\bibitem{MantegnaStanley} Mantegna R.N., Stanley H.E., \emph{Introduction to Econophysics: Correlations and Complexity in Finance}. Cambridge University Press, Cambridge (1999).


\bibitem{McCauley} J.L. McCauley, \emph{Dynamics of Markets
Econophysics and Finance}, 2007

\bibitem{Bouchaud} J-P. Bouchaud, M. Potters, 
 \emph{Theory of Financial Risk and Derivative Pricing: From Statistical Physics to Risk Management}, Cambridge University Press, 2003. 

\bibitem{Bass} T. A.  Bass,  \emph{The Predictors: How a Band of Maverick Physicists Used Chaos Theory to Trade Their Way to a Fortune on Wall Street}, Henry Holt \& company New York, 2000. 

\bibitem{Sornette-libro} D. Sornette,  \emph{Why Stock Markets Crash:
Critical Events in Complex Financial Systems} 2004

\bibitem{Galam} S. Galam, \emph{Sociophysics
A Physicist's Modeling of Psycho-political Phenomena}
2012, XXIII

\bibitem{Stauffer} D. Stauffer,  \emph{Opinion Dynamics and Sociophysics} Encyclopedia of Complexity and Systems Science
2009, pp 6380-6388

\bibitem{Helbing1}
D.  Helbing,   \emph{ Quantitative Sociodynamics}, Kluwer, Dordrecht (1995)

\bibitem{Castellano}
Castellano C., Fortunato S. and Loreto V., \emph{ Statistical physics of social dynamics}, Reviews of Modern Physics 81: 591-646 (2009)

\bibitem{hart}
Hart, N., \emph{Alfred Marshall and Modern Economics}, 2013 (Palgrave MacMillan, London), original citation at p.16.

\bibitem{simon} 
Simon, H.A.,  \emph{Models of Man}, 1957 (Wiley, New York).

\bibitem{friedman} 
Friedman, M., \emph{A Theory of the Consumption Function}, 1956 (Princeton University Press, Princeton, N.J.).


\bibitem{MC} 
Metropolis, Nicholas, Rosenbluth, Arianna W., Rosenbluth, Marshall N., Teller. Augusta H.. and Teller, Edward.“Equation of State Calculations by Fast Computing Machines.” Journal of Chemical Physics 21 (1953) 1087.

\bibitem{Binder} Binder, K., Heermann, D.W.: Monte Carlo Simulation in Statistical Physics: An Introduction. Springer, Berlin (1988)

\bibitem{Benzi} Benzi R., Parisi G., Sutera A. and Vulpiani A., \emph{Stochastic resonance in climate change}, Tellus, 34, 10-16 (1982).

\bibitem{Nicolis} Nicolis, C., \emph{Stochastic aspects of climatic transitions-response to a periodic forcing}, Tellus 34, 1-9 (1982).

\bibitem{Gammaitoni} Gammaitoni, L., Hanggi, P., Jung, P., Marchesoni, F.: Rev. Mod. Phys. 70, 1 (1998) 

\bibitem{Thurner1} D. Stokic, R. Hanel, and S. Thurner, Phys. Rev. E 77, 061917 (2008).
\bibitem{Thurner2}  R. Hanel,M. P\"{o}chacker, M. Sch\"{o}lling, and S. Thurner, Plos One 7, e36679 (2012).

\bibitem{Caruso1} F. Caruso, S. F. Huelga, and M. B. Plenio, Phys. Rev. Lett. 105,
190501 (2010).

\bibitem{Mossa} 
Mossa F., Wardb L. M., Sannitac W. G., \emph{Stochastic resonance and sensory information processing: a tutorial and review of application}, Clinical Neurophysiology 115, 267–281 (2004).


\bibitem{Peter}
L.J. Peter, R. Hull,    \emph{ The Peter Principle: Why Things Always Go Wrong}, William Morrow and Company, New York, 1969.

\bibitem{Pluchino1}
A. Pluchino, A. Rapisarda and C. Garofalo, The Peter Principle revisited:  a computational study, Physica A, 389, 467 (2010).  This paper  was quoted by several blogs and specialized newspapers, among which the MIT blog, the New York Times and the Financial Times, and it was also awarded the IG Nobel prize 2010 for "Management" . For more info see the webpage: www.pluchino.it/ignobel.html  


\bibitem{Pluchino2}
A. Pluchino, A. Rapisarda and C. Garofalo, Efficient promotion strategies in hierarchical organizations, Physica A, 390   3496 (2011)


\bibitem{Pluchino3}
A. Pluchino, C. Garofalo, A. Rapisarda, S. Spagano, M. Caserta, Accidental Politicians: How Randomly Selected Legislators, Can improve  Parliament Efficiency, Physica A , 390, 3944 (2011).     
See also http://www.pluchino.it/Parliament.html


\bibitem{Sornette2} J.B. Satinover and D. Sornette, ÔøΩIllusion of controlÔøΩ in Time-Horizon Minority and Parrondo Games, Eur. Phys. J. B 60, 369ÔøΩ384 (2007); J.B. Satinover  e D. Sornette, ÔøΩIllusory versus genuine control in agent-based gamesÔøΩ,  Eur. Phys. J. B 67, 357ÔøΩ367 (2009)

\bibitem{Wiseman} 
R. Wiseman, \emph{Quirkology}. Macmillan, London (2007)

\bibitem{Porter}  Porter G. E., 
\emph{The long term value of analysts advice in the Wall Street
Journals investment dartboard contest},
J. Appl. Finance, 14, 720 (2004). 

\bibitem{Cass}  A. Clare, N. Motson, S. Thomas,  
\emph{An evaluation of alternative equity indices}.
Cass consulting report March 2013, http://www.cassknowledge.com/research/article/evaluation-alternative-equity-indices-cass-knowledge

\bibitem{Biondo1} A.E. Biondo, A. Pluchino, A. Rapisarda, Journal of Statistical Physics 151 (2013) 607.

\bibitem{Biondo2} A.E. Biondo, A. Pluchino, A. Rapisarda, D. Helbing, (2013) PLOS ONE 8(7): e68344. doi:10.1371/journal.pone.0068344

\bibitem{Biondo3}  A.E. Biondo, A. Pluchino, A. Rapisarda, D. Helbing, Phys. Rev. E 88, 062814 (2013)

\bibitem{smith}
Smith, A., \emph{An Inquiry Into The Nature and Causes of The Wealth of Nations}, 1904 (Meuthen \& Co., London). 

\bibitem{ricardo}
Ricardo, D., \emph{Principles of Political Economy and Taxation}, 1817 (John Murray, London).

\bibitem{malthus}
Malthus, T.R., \emph{Principles of Political Economy}, 1836 (W. Pickering, London).

\bibitem{marshall} 
Marshall A., \emph{Principles of Economics}, 1920 (MacMillan, London).

\bibitem{edgeworth}
Edgeworth, F.Y., \emph{Mathematical Psychics: An Essay on the Application of Mathematics to the Moral Sciences}, 1881 (C. Kegan Paul \& Co., London)

\bibitem{jevons}
Jevons, W.S., \emph{The Theory of Political Economy}, 1888 (MacMillan, London).

\bibitem{walras}
Walras, M.E.L., \emph{Elements D'Economie Politique Pure}, 1926 (R.Pichon et R.Durand-Auzias Editeurs, Paris).

\bibitem{bohm-bawerk}
von Bohm-Bawerk, E., \emph{The Positive Theory of Capital}, 1889 (MacMillan, London).

\bibitem{menger}
Menger, C., \emph{Principles of Economics}, 1871 - 2007 (Ludwig von Mises Institute, USA).

\bibitem{fisher}
Fisher, I., \emph{The Purchasing Power of Money}, 1922 (MacMillan NY).

\bibitem{pareto}
Pareto, V., \emph{Manuale di Economia Politica}, 1906 (Società Editrice Libraria, Milano).

\bibitem{keynes}
Keynes, J.M., \emph{The General Theory of Unemployment, Interest and Money}, 1936 (MacMillan, London).

\bibitem{arrow-nerlove}
Arrow, K.J. and Nerlove M., A Note on Expectations and Stability. \emph{Econometrica}, 1958, {\bf 26}, 297-305.

\bibitem{friedman2} 
Friedman, M.,The Role of Monetary Policy, \emph{The American Economic Review}, 1968, 1-17.

\bibitem{phelps}
Phelps, E., Phillips Curve Expectations of Inflation, and Output Unemployment Over Time, \emph{Economica}, 1967, {\bf 34} (135), 254-281.

\bibitem{cagan}
Cagan, P., \emph{The Monetary Dynamics of Hyperinflation}, In Friedman M., (ed.) \emph{Studies in the Quantity Theory of Money}, 1956 (University of Chicago Press, Chicago).

\bibitem{muth}
Muth, J.F., Rational Expectation and the Theory of Price Movements. \emph{Econometrica}, 1961, {\bf 29}, 315-335.

\bibitem{lucas}
Lucas, R.E., Expectations and the Neutrality of Money. \emph{Journal of Economic Theory}, 1972, {\bf 4}, 103-124.

\bibitem{sargent-wallace}
Sargent, T.J. and Wallace N., Rational Expectations, the Optimal Monetary Instrument, and the Optimal Money Supply Rule. \emph{Journal of Political Economy}, 1975, {\bf 83} (2), 241-254.

\bibitem{fama}
Fama, E.F., Efficient Capital Markets: a Review of Theory and Empirical Work. \emph{Journal of Finance}, 1970,  {\bf 25}, 383-423.

\bibitem{jensen}
Jensen, M., Some anomalous evidence regarding market efficiency. \emph{Journal of Financial Economics}, 1978,  {\bf 6}, 95-101.

\bibitem{malkiel}
Malkiel, B., \emph{Efficient market hypothesis}. New Palgrave Dictionary of Money and Finance, 1992 (Macmillan, London).

\bibitem{cutler}
Cutler, D.M., Poterba, J.M., Summers, L.H.,  What moves stock prices? \emph{Journal of Portfolio Management}, 1989, April, 4-12.

\bibitem{engle}
Engle, R., Autoregressive conditional heteroscedasticity with estimates of the variance of UK inflation, \emph{Econometrica}, 1982, {\bf 50}, 987-1008.

\bibitem{mandelbrot} 
Mandelbrot, B.B., The variation of certain speculative prices. \emph{Journal of Business}, 1963,  {\bf 36}, 394-419.

\bibitem{mandelbrot2} 
Mandelbrot, B.B., \emph{Fractals and Scaling in Finance}, 1997, (Springer, New York).

\bibitem{lux2}
Lux T., The stable Paretian hypothesis and the frequency of large returns: an examination of major German stocks. \emph{Applied Financial Economics}, Vol. 6, pp. 463-475 (1996).


\bibitem{brock}
Brock W.A., Pathways to Randomness in the Economy: Emergent Non-Linearity and Chaos in Economics and Finance.\emph{Estudios Econ\'omicos}, Vol. 8, pp. 3-55 (1993).

\bibitem{brock2}
Brock W.A., \emph{Asset Prices Behavior in Complex Environments}. In: Arthur W.B., Durlauf S.N., and Lane D.A., eds., \emph{The Economy as an Evolving Complex System II}, Addison-Wesley, Reading, MA, pp. 385-423 (1997).

\bibitem{brock-hommes}
Brock W.A., and Hommes C.H., A Rational Route to Randomness. \emph{Econometrica}, Vol. 65, pp.1059-1095 (1997).

\bibitem{chiarella}
Chiarella C., The Dinamics of Speculative Behavior. \emph{Annals of Operations Research}, Vol. 37, pp. 101-123 (1992).

\bibitem{chiarella-he}
Chiarella C., and He T., Heterogeneous Beliefs, Risk and Learning in a Simple Asset Pricing Model, \emph{Computational Economics - Special issue: Evolutionary processes in economics}, Vol. 19 (1), pp. 95-132 (2002). 

\bibitem{degrauwe}
DeGrauwe P., DeWachter H., and Embrechts M., \emph{Exchange Rate Theory. Chaotic Models of Foreign Exchange Markets, Blackwell} (1993).

\bibitem{frankel-froot}
Frankel J.A., and Froot K.A., Chartists, Fundamentalists and the Demand for Dollars. \emph{Greek Economic Review}, Vol. 10, pp. 49-102 (1988).

\bibitem{lux}
Lux T., Herd Behavior, Bubbles and Crashes. \emph{The Economic Journal}, Vol. 105, pp. 881-896.

\bibitem{wang}
Wang J., A Model of Competitive Stock Trading Volume. \emph{Journal of Political Economy}, Vol. 102, pp. 127-168 (1994).

\bibitem{zeeman}
Zeeman E.C., The Unstable Behavior of Stock Exchange. \emph{Journal of Mathematical Economics}, Vol. 1, pp. 39-49 (1974).

\bibitem{gabaix} 
Gabaix X., Gopikrishnan P., Plerou V., Stanley H.E., 
\emph{A theory of power-law distributions in financial market fluctuations}, 
Nature 423: 267-72 (2003) 

\bibitem{preis} 
T. Preis, D. Y. Kenett, H. E. Stanley, D. Helbing, E. Ben-Jacob,  
\emph{Quantifying the behavior of stock correlations under market stress} 
Sci. Rep. 2, 752; 10.1038/srep00752 (2012).

\bibitem{Carbone} 
Carbone A., Castelli G., Stanley H. E., (2004) Time dependent  Hurst exponent in financial time series. Physica A 344:  267-271.

\bibitem{murphy} 
Murphy J.J., \emph{Technical Analysis of the Financial Markets: A Comprehensive Guide to Trading Methods and Applications}, New York Institute of Finance (1999) 

\bibitem{Taleb} Taleb N.N., \emph{Fooled by Randomness: The Hidden Role of Chance in the Markets and in Life}. Random House, NY (2005).

\bibitem{Buchanan} 
M.Buchanan, 
{\it Forecast: What Physics, Meteorology, and the Natural Sciences Can Teach Us About Economics}, 
(Bloomsbury, 2013)

\bibitem{Helbing-Kern}
D. Helbing and D. Kern, 
\emph{Non-equilibrium price theories} 
Physica A 287, 259–268 (2000).

\bibitem{HelbingVicsek} 
D. Helbing, I. Farkas, T. Vicsek,
\emph{Simulating dynamical features of escape panic},
Nature 407, 487-490 (2000).

\bibitem{PeinkeNature}
S. Ghashghaie, W. Breymann, J. Peinke, P. Talkner, Y. Dodge,
Nature 381, 767 - 770 (1996). 

\bibitem{Farmer1}
J.D. Farmer, Ind. and Corp. Change, 11 (5): 895 (2002).
doi: 10.1093/icc/11.5.895

\bibitem{Bouchaud2}
J-P. Bouchaud, M. Potters, {\it Theory of financial risks: from statistical physics to risk management}, (2004) Lavoisier

\bibitem{Sornette}
D. Sornette, {\em Why stock markets crash: critical events in complex financial systems} (Princeton University Press, 2003).

\bibitem{Helbing-Nature}
D. Helbing, 
\emph{Globally networked risks and how to respond},
Nature 497, 51-59 (2013)

\bibitem{krawiecki}
A. Krawiecki, J. A. Holyst, and D. Helbing,  
\emph{Volatility clustering and scaling for financial time series due to attractor bubbling} 
Phys. Rev. Lett. 89, 158701 (2002).

\bibitem{parisi}
D. R. Parisi, D. Sornette, D. Helbing 
\emph{Financial price dynamics and pedestrian counterflows: A comparison of statistical stylized facts} 
Phys. Rev. E 87, 012804 (2013). 

\bibitem{Bikhchandani-Hirshleifer-Welch} 
S. Bikhchandani, D. Hirshleifer, and I. Welch, \textit{Informational cascades and rational herding: An annotated bibliography and resource reference, working paper} (Anderson School of Management, UCLA, available at www.info-cascades.info, 2008)

\bibitem{Bak}
P. Bak, C. Tang, K. Wiesenfeld, Phys. Rev. Lett. 59, 381 (1987).

\bibitem{Olami} 
Z. Olami, H. J. S. Feder, and K. Christensen, Phys. Rev. Lett. 68, 1244 (1992).

\bibitem{Caruso}
F. Caruso, A. Pluchino, V. Latora, S. Vinciguerra, A. Rapisarda, 
\emph{Analysis of self-organized criticality in the OFC model and in real earthquakes} , 
Phys. Rev. E 75, 055101(R) (2007). 

\bibitem{Mantegna1}
F. Lillo and R. N. Mantegna, 
\emph{Power-law relaxation in a complex system: Omori law after a financial market crash},
Phys. Rev. E 68, 016119 (2003).

\bibitem{Caruso3}
F. Caruso, V. Latora, A. Pluchino, A. Rapisarda, B. Tadic, Eur. Phys. J. {\bf B 50 }, 243  (2006).

\bibitem{barabasi}
A.L. Barabasi, R. Albert, Science 286, 509 (1999)

\bibitem{nota}
Please note that the final distribution of the capital of random traders obtained here differs from the exponential one found in ref.\cite{Biondo3} because there we did not endow all the agents with exactly the same amounts of 1000 credits, but we considered an initial Gaussian distribution with an average value of 1000 credits.

\bibitem{Economics2.0}
D. Helbing, 
\emph{Economics 2.0: The Natural step towards a self-regulating, participatory market society}, 
Evol. Inst. Econ. Rev. 10(1): 3–41 (2013)


\end{thebibliography}
\end{document}